\documentclass[10pt]{article}

\usepackage{newtxtext,newtxmath}
\usepackage{graphicx}
\graphicspath{{FigK/}}  % Add both search paths

\usepackage[letterpaper,margin=1in]{geometry}

\renewenvironment{abstract}
	{\quotation}
	{\endquotation}

\makeatletter
\renewcommand{\fnum@figure}{\textbf{Figure \thefigure}}
\renewcommand{\fnum@table}{\textbf{Table \thetable}}
\makeatother

\usepackage{scicite}

\usepackage{url}

\newcommand{\mc}[1]{\ensuremath{\mathcal{#1}}}   % For Hilbert spaces and subspaces
\newcommand{\mb}[1]{\ensuremath{\mathrm{#1}}}   % For Hamiltonians
\newcommand{\ms}[1]{\ensuremath{\mathsf{#1}}} % For subscripts or driver labels (already defined)
\newcommand{\mt}[1]{\ensuremath{\mathtt{#1}}}
\newcommand{\abbr}[1]{\ensuremath{\mathtt{#1}}}  % For abbreviations like \LM, \GM, etc.

\newcommand{\XX}{\ms{XX}}
\newcommand{\GIC}{\mt{GIC}}
\newcommand{\MIS}{\mt{MIS}}

\renewcommand{\bar}{\overline}

\newcommand{\DDD}{{\sc Dic-Dac-Doa{}}}

\newcommand{\ZZ}{\ms{ZZ}}

\newcommand{\x}[1]{\mt{x}(#1)}
\newcommand{\jxx}[1]{\mt{jxx}(#1)}

\newcommand{\shz}[1]{\tilde{\sigma}^z_{#1}}

\newcommand{\ver}{{\ms{V}}}
\newcommand{\edge}{{\ms{E}}}

\newcommand{\Jxx}{J_{\ms{xx}}}
\newcommand{\Jzz}{J_{\ms{zz}}}

\newcommand{\LM}{\abbr{LM}}
\newcommand{\GM}{\abbr{GM}}

\newcommand{\MIC}{\abbr{dMIC}}
\newcommand{\MDC}{\abbr{dMDC}}

\newcommand{\Gdis}{\ensuremath{G_{\ms{dis}}}}
\newcommand{\Gshare}{\ensuremath{G_{\ms{share}}}}

\newcommand{\cl}{\ms{Clique}}
\newcommand{\Heff}{\mb{H}_1^{\ms{eff}}}

\newcommand{\Hinter}{\mb{H}_{\ms{inter\!-\!block}}}
\newcommand{\weff}{w^{\text{\tiny eff}}}

\newcommand{\Jxxlift}{\Jxx^{\text{\tiny lift}}}
\newcommand{\Jxxsteer}{\Jxx^{\text{\tiny steer}}}
\newcommand{\Jxxsep}{\Jxx^{\text{\tiny sep}}}
\newcommand{\Jxxsink}{\Jxx^{\text{\tiny sink}}}
\newcommand{\Jzzinter}{\Jzz^{\text{\tiny inter}}}

\newcommand{\SLE}{\mathcal{L}^{\text{ind}}}
\newcommand{\Hcore}{\mb{H}_{\ms{core}}}

\def\scititle{
  Exponential Quantum Speedup on Structured Hard Instances of Maximum Independent Set
}
\title{\bfseries \boldmath \scititle}

\author{
  V.~Choi\\
   Gladiolus Veritatis Consulting Co.\footnote{{https://www.vc-gladius.com}}
}

\begin{document} 

% Insert the title and author list
\maketitle

\begin{abstract} \bfseries \boldmath
  Establishing quantum speedup for computationally hard problems of practical relevance, particularly combinatorial optimization problems, remains a central challenge in quantum computation.
In this work, we identify a structurally defined family of classically hard maximum independent set (MIS) instances,
and design and analyze a non-stoquastic adiabatic quantum optimization algorithm that exploits this structure.
The algorithm runs in polynomial time and achieves an exponential speedup over both transverse-field quantum annealing and state-of-the-art classical solvers on these instances,
under assumptions supported by analytical and numerical evidence.
We identify the essential quantum mechanism enabling the speedup as the use of a non-stoquastic $\XX$-driver to
access a larger sign-structured admissible subspace beyond the stoquastic regime, which allows
sign-generating quantum interference to create smooth evolution paths that bypass tunneling.
This identifies a distinctive quantum mechanism underlying the speedup and
explains why no efficient classical analogue is likely to exist.
In addition, our analysis produces scalable small-scale models, derived
from our structural reduction, that capture the essential dynamics of
the algorithm. These models provide a concrete opportunity for
verification of the quantum advantage mechanism on currently available
universal quantum computers.
\end{abstract}

\section{Introduction}
\noindent
Breakthrough work by Shor~\cite{Shor1997} demonstrated the potential of quantum computation to achieve exponential speedup for problems believed to be classically intractable.
One of the broad attractions of quantum computation is its potential to address NP-hard combinatorial optimization problems, which are widely believed to be intractable for classical computers.
However, since Shor's result, provable and practically relevant quantum speedups have predominantly been realized in quantum simulation; see, e.g., the references in~\cite{Babbush2025GrandChallenge}.
Despite substantial effort, much of the work on quantum optimization has emphasized heuristic or hybrid approaches rather than rigorous speedup guarantees.
As a result, provably advantageous quantum algorithms for practically relevant optimization problems have remained elusive; see, e.g., \cite{Dupont2023SciAdv,Jordan2025DQI}.

In this work, we identify a structurally defined family of classically hard maximum independent set (MIS) instances, referred to as \GIC{} graphs.
We design and analyze a non-stoquastic adiabatic quantum optimization algorithm, called \DDD{}\footnote{The acronym originally stood for \textit{Driver graph from Independent Cliques, Double Anti-Crossing, Diabatic quantum Optimization Annealing}.}, first proposed in 2021~\cite{Choi2021}.
The algorithm runs in polynomial time and achieves exponential speedup over both transverse-field quantum annealing (TFQA)
and classical algorithms on \GIC{} graphs, under assumptions supported by analytical and numerical evidence.
The underlying structural principles of the graphs are not contrived, but capture intrinsic sources of hardness in MIS,
and can be extended to broader classes of instances, including those relevant to real-world applications.
This contribution directly corresponds to two central stages emphasized by Babbush \emph{et al.}:
``the identification of concrete problem instances expected to exhibit quantum advantage (Stage~II) and the connection of such instances to real-world use cases (Stage~III)''~\cite{Babbush2025GrandChallenge}.

The MIS problem is a canonical NP-hard combinatorial optimization problem~\cite{GareyJohnson1979}
and has served as a central benchmark for quantum optimization.
Indeed, quantum hardware has been designed specifically for solving MIS problems~\cite{Choi2008,Choi2011,Johnson2011,Ebadi2022ScienceMIS}.

Adiabatic quantum computing (AQC) was originally proposed as a general-purpose framework for solving optimization problems~\cite{Farhi2000,Farhi2001,Nishimori2001SpinGlasses}.
It was later shown to be polynomially equivalent to the standard circuit model of quantum computation~\cite{Aharonov2007AQCEq}.
Implementing an adiabatic quantum algorithm on a gate-based quantum computer is referred to as digitized AQC~\cite{Barends2016NatureDAQC}.
In AQC, the system is initialized in the ground state of a simple Hamiltonian and slowly evolved toward a final Hamiltonian encoding the problem of interest.
While this analog paradigm is conceptually simple from an algorithmic design perspective, rigorous analysis of its computational performance,
in particular its runtime scaling, has proven challenging.
In practice, adiabatic quantum algorithm (or quantum annealing) has most often been treated as a heuristic, black-box optimization paradigm,
in which little problem-specific structure is incorporated into the algorithmic design, and performance is typically assessed empirically rather than through analytical characterization.

Here, we show how to explicitly incorporate problem structure into the design and analysis of a quantum optimization algorithm,
and develop an analytical framework for characterizing its performance.
Our approach departs from the traditional spectral-gap perspective~\cite{AQC-Review} and instead measures the efficiency of the algorithm by the presence or absence of an anti-crossing (see Section~5 of Supplement for a precise definition).
The key idea is to infer this behavior directly from the crossing properties of bare energy levels of relevant subsystems,
without explicitly constructing the effective two-level Hamiltonian.

The \DDD{} algorithm modifies standard TFQA by adding a specially designed non-stoquastic $\XX$-driver term,
aimed at overcoming the small-gap anti-crossings that plague TFQA on \GIC{} graphs.
Despite their structured construction, we argue that \GIC{} graphs exhibit classical hardness:
solving MIS on such instances would require exponential time unless \( \mathrm{P} = \mathrm{NP} \).
We provide numerical results using state-of-the-art classical solvers on representative instances.
We further conjecture that no tailored classical algorithm can solve these structured instances efficiently.

This raises a natural question: what quantum resource enables \DDD{} to succeed on instances
that remain intractable for both TFQA and classical algorithms?
As also emphasized in~\cite{Barends2016NatureDAQC}, non-stoquastic Hamiltonians may play an essential role in AQC,
both for universality and for improved performance on hard optimization problems.

Here, we explicitly identify the essence of going beyond stoquasticity:
the ability to exploit a larger sign-structured subspace, which enables new evolution paths that are dynamically inaccessible in the stoquastic regime.
This mechanism is fundamentally different from the conventional view that quantum speedup in Hamiltonian-based optimization arises from quantum tunneling.
This perspective is rooted in a fundamental feature of quantum mechanics:
the wavefunction of a quantum state, expressed in a given basis, can have both positive and negative amplitudes,
in contrast to classical probabilities, which must be non-negative.
We therefore introduce terminology to characterize sign structure.
A quantum state expressed in a given basis is said to be \emph{same-sign} if all amplitudes in that basis are non-negative,
and \emph{opposite-sign} if it contains both positive and negative amplitudes; see Section~1 of Supplement for a precise definition.
Correspondingly, a basis is called same-sign if it consists entirely of same-sign states, and opposite-sign otherwise.
Subspaces/sectors and matrix blocks are classified according to the sign structure of the basis in which they are expressed.

For stoquastic Hamiltonians, Perron--Frobenius theorem implies that the ground state is necessarily same-sign in the computational basis.
As a result, the admissible subspace dynamically accessible to the ground state remains confined to the same-sign sector.
By contrast, introducing a non-stoquastic $\XX$ driver opens access to opposite-sign sectors, enlarging the admissible subspace.
Within this expanded subspace, \emph{sign-generating quantum interference}---interference that produces negative amplitudes in the computational basis---enables a \emph{smooth evolution path} that bypasses tunneling.
This identifies the distinctive quantum mechanism underlying the speedup and explains why no efficient classical analogue is likely to exist.

 %995

\section{Results}
\subsection{\GIC{} Graph Instances}

We consider a structurally defined family of graphs, referred to as \GIC{} graphs.
Each such graph encodes a planted maximum independent set (\MIS{}) together with an extensive collection of competing maximal independent sets.
The construction organizes vertices into clique-based blocks, with inter-block couplings chosen to preserve the planted MIS while inducing classical hardness.
Each clique-based block corresponds to a \MIC{} as defined in \cite{Beyond}; see Section~3.2 of Supplement.

\begin{figure}
  \centering
  \begin{tabular}{cc}
    \includegraphics[width=0.15\textwidth]{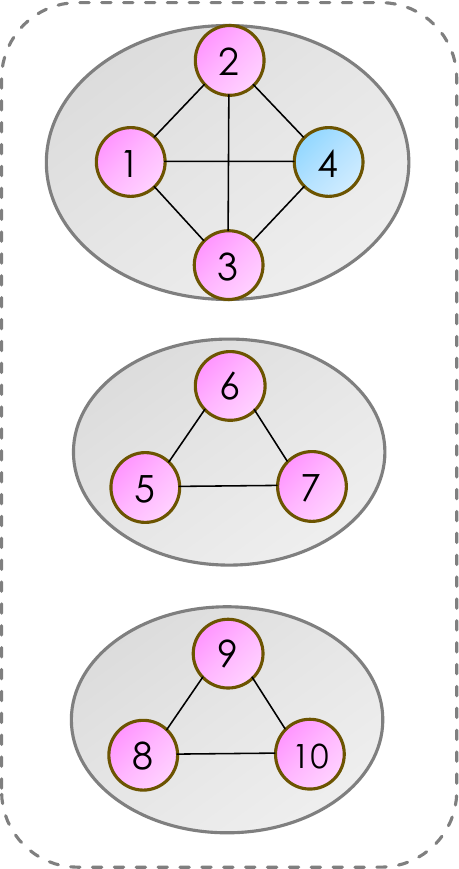} &
    \includegraphics[width=0.45\textwidth]{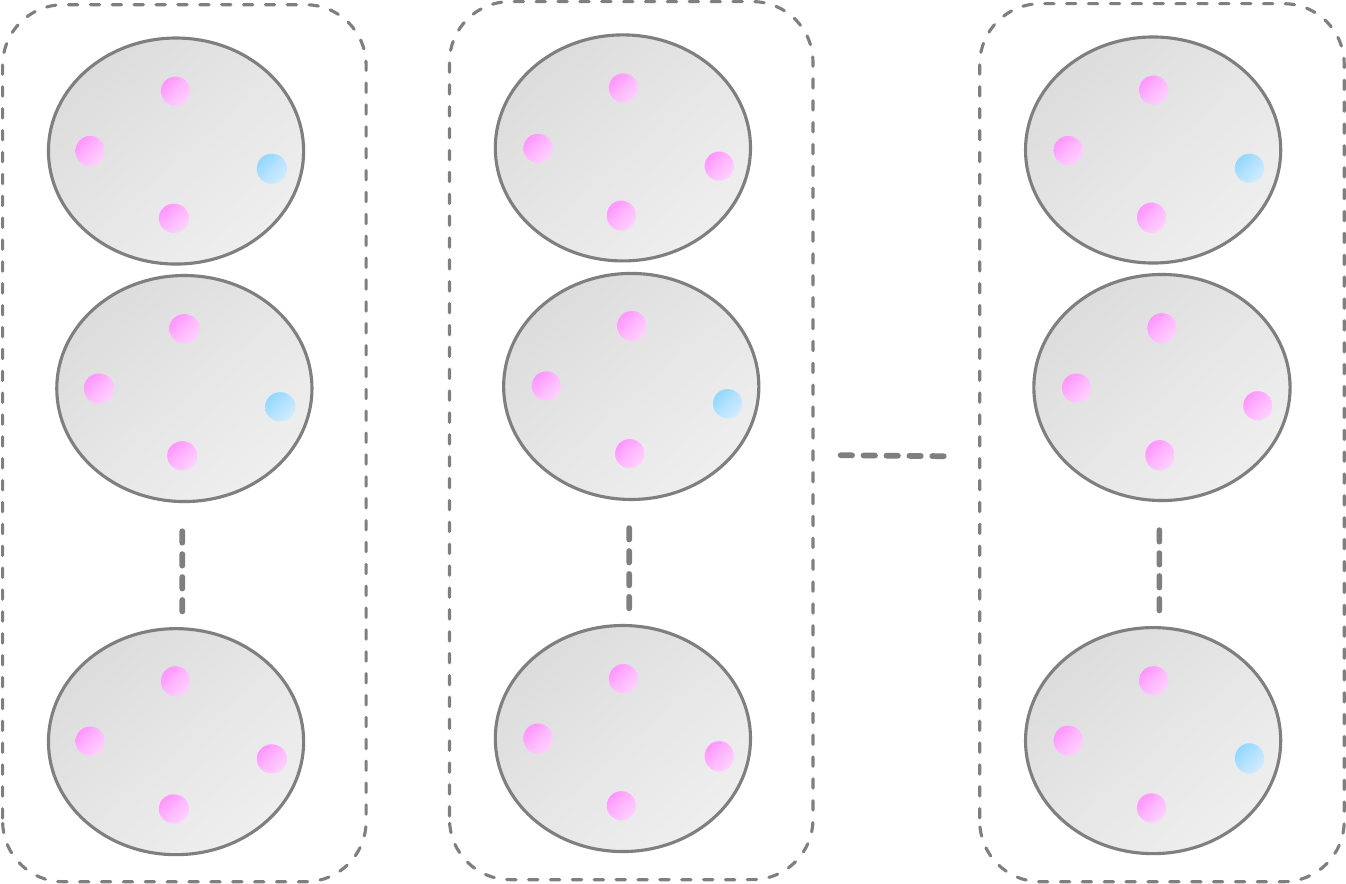} \\
    (\textbf{a}) & (\textbf{b})
  \end{tabular}
  \caption{\textbf{Structure of a \MIC{} block and a \GIC{} graph.}
  (\textbf{a}) A single \MIC{} block consisting of three cliques of sizes $(4,3,3)$.
  Each maximal independent set is obtained by selecting exactly one vertex from each clique,
  yielding $4 \times 3 \times 3$ degenerate maximal independent sets.
  The highlighted vertex (blue) denotes the planted MIS vertex; all remaining vertices are non-MIS vertices.
  (\textbf{b}) Schematic structure of a \GIC{} graph composed of multiple blocks.
  Edges within cliques and between blocks are omitted for clarity.
  Blocks are initially connected by complete bipartite couplings, after which cross-block edges are selectively removed to embed a unique planted global MIS spanning multiple blocks, while preserving degree-oblivious local structure.}
  \label{fig:GIC}
\end{figure}

A \MIC{} of size \( k \) consists of \( k \) \emph{independent cliques} (i.e., no edges exist between them),
denoted as \( \text{Clique}(n_i) \), where \( n_i \) is the clique size, for \( 1 \le i \le k \).
Each maximal independent set within a \MIC{} is formed by selecting exactly one vertex from each clique.
The resulting degeneracy is therefore
\(
\prod_{i=1}^{k} n_i .
\)
We will refer to each such \MIC{} as a \emph{\MIC{} block}.
An illustration of a single \MIC{} block is shown in Fig.~\ref{fig:GIC}(\textbf{A}).

A \GIC{} instance consists of at least three such \MIC{} blocks.
A schematic illustration of the full construction is shown in Fig.~\ref{fig:GIC}(\textbf{B}).
Within each block, vertices belonging to the same clique are fully connected, and different blocks are initially connected via complete bipartite graphs.
A unique global MIS is then planted by selecting a fixed number of cliques in each block and choosing one representative vertex per selected clique.
All cross-block edges among these representatives are removed, ensuring that the planted set is independent and maximal while necessarily spanning at least three blocks.

To prevent planted MIS vertices from being distinguishable by degree while preserving global optimality, the graph is further modified using an \emph{anchored degree thinning} procedure with global uniqueness protection.
Because this procedure depends only on local neighborhood information, it can be implemented efficiently.
Overall, the \GIC{} instances considered here are constructible in polynomial time.

\paragraph{Correspondence between \MIC{} and degenerate local minima.}
The MIS problem can be encoded in the ground state of the MIS-Ising problem Hamiltonian, as detailed in
Section~2 of Supplement.
Each maximal independent set corresponds to a local minimum of the problem Hamiltonian.
A collection of maximal independent sets all having the same size \( m \)
corresponds to a set of \emph{degenerate} local minima with equal energy \( -m \).
Thus, each \MIC{} induces a set of degenerate local minima (\LM{})---forming a wide basin---in the MIS-Ising energy landscape.
The terms \LM{} and \MIC{} are thus used interchangeably when no ambiguity arises,
with \LM{} used in place of \MIC{} when emphasizing the corresponding energy landscape structure.
We use \GM{} to refer both to the (global) maximum independent set and to its corresponding global minimum in the energy landscape.

Further motivation for the \GIC{} construction and its relation to
the structural origin of classical hardness in MIS
is discussed in Section~3.2 of Supplement.

\paragraph{Anchored degree thinning.}
[Move to Supplement.]
To prevent planted MIS vertices from being distinguishable by degree while preserving global optimality,
we apply an \emph{anchored degree thinning} procedure.
For each non-MIS vertex $v$ belonging to a clique that contains a planted MIS vertex $m$,
incident edges are selectively removed so that the post-thinning degree of $v$
is the same as the degree of its anchor $m$.
Edge removals are restricted to neighbors that
(i) are not part of the planted MIS,
(ii) lie outside the clique of $v$, and
(iii) are neighbors of the anchor $m$.
A local safeguard further ensures that the set of removed neighbors does not form
an independent set, which guarantees that replacing $m$ by $v$ cannot yield a maximal
independent set of equal size, thereby preserving the global uniqueness of the planted MIS.
Each vertex is allowed to participate in thinning at most once.
After a vertex has been thinned, it is excluded from further thinning operations,
including participation as a removable neighbor in subsequent thinning steps.
This prevents successive thinning steps from propagating across the graph and
ensures that degree reductions do not concatenate to form large independent sets.
Because the procedure depends only on local neighborhood information and clique membership,
it can be implemented efficiently and does not alter the planted solution.

\subsection{Classical Hardness}

We evaluate classical performance using the representative algorithm (ReduMIS) from KaMIS~\cite{Lamm2017ReduMIS,Lamm2016KaMIS}, a state-of-the-art framework for the maximum independent set problem that combines aggressive graph reductions with evolutionary local search and has been shown to outperform exact solvers on a wide range of benchmark instances.
We further include a parallel tempering (PT) baseline~\cite{Hukushima1996PT}, which represents a standard stochastic optimization approach that probes the energy landscape directly.

The goal of this evaluation is to establish that leading classical approaches encounter difficulty due to the underlying combinatorial structure of the instances, as expected.

In our experiments, we restrict attention to the simplest nontrivial instances consisting of exactly three \MIC{} blocks with uniform clique sizes within each block.
The description of one example \GIC{} family and the corresponding solver results are presented in Table~\ref{tab:gic-classical}.

\begin{table}[t]
\centering
\caption{\textbf{Classical solver performance on a scaled \GIC{} family.}
(\textbf{Top}) Description of an example \GIC{} family parametrized by an integer $k$.
Each instance consists of three blocks specified by $(m_i,n_i)$, $i=1,2,3$, where for each block $i$, $m_i$ denotes the number of cliques and each clique has size $n_i$.
The planted MIS is specified by parameters $(g_1,g_2,g_3)$, where $g_i$ denotes the number of MIS vertices in block $i$ (one vertex per selected clique).
The planted MIS size is $|\MIS{}| = g_1 + g_2 + g_3$.
(\textbf{Bottom}) Classical solver performance for selected instances from this family.
As $k$ increases, ReduMIS transitions from early termination to cutoff-limited success; for the largest instance shown, the time limit was increased from 1200 s to 2400 s to allow recovery of the planted MIS.
(Reported ReduMIS runtimes may slightly exceed the nominal cutoff due to solver overhead.)
Parallel tempering (PT), run with a fixed 60 s time limit, becomes trapped in suboptimal basins and fails to reach the planted MIS across all instances.
All experiments were performed on a local Apple MacBook Pro equipped with an Apple M1~Max processor and 64~GB of memory.}
\label{tab:gic-classical}

% ---------- Family definitions ----------
%\textbf{Instance families}
\vspace{0.5em}

\textbf{Instance definition (\GIC{} family)}
\vspace{0.5em}

\begin{tabular}{c c c c c}
\hline
 &
$(m_1,n_1)$ &
$(m_2,n_2)$ &
$(m_3,n_3)$ &
$(g_1,g_2,g_3)$ \\
\hline
 & $(4+4k,9)$ & $(8+4k,6)$ & $(3+2k,4)$ & $(2+2k,\,6+2k,\,2)$ \\ 
\hline
\end{tabular}

\vspace{0.8em}
\textbf{Classical solver performance}

\begin{tabular}{c c c c c}
\hline
$k$ & $|V|$ & Planted MIS & ReduMIS runtime / time limit (s) & PT output (time limit = 60 s)
\\
\hline
0 & 96  & 10 & 50.2 / 1200   & 8  \\
3 & 300 & 22 & 414.0 / 1200  & 20 \\
5 & 436 & 30 & 798.4 / 1200  & 24 \\
7 & 572 & 38 & 2676.1 / 2400 & 36 \\
\hline
\end{tabular}
\end{table}

\subsection{\DDD{} on \GIC{} Graphs}

\begin{figure}[!htbp]

   \begin{tabular}{ccc}
     {\includegraphics[width=0.30\textwidth]{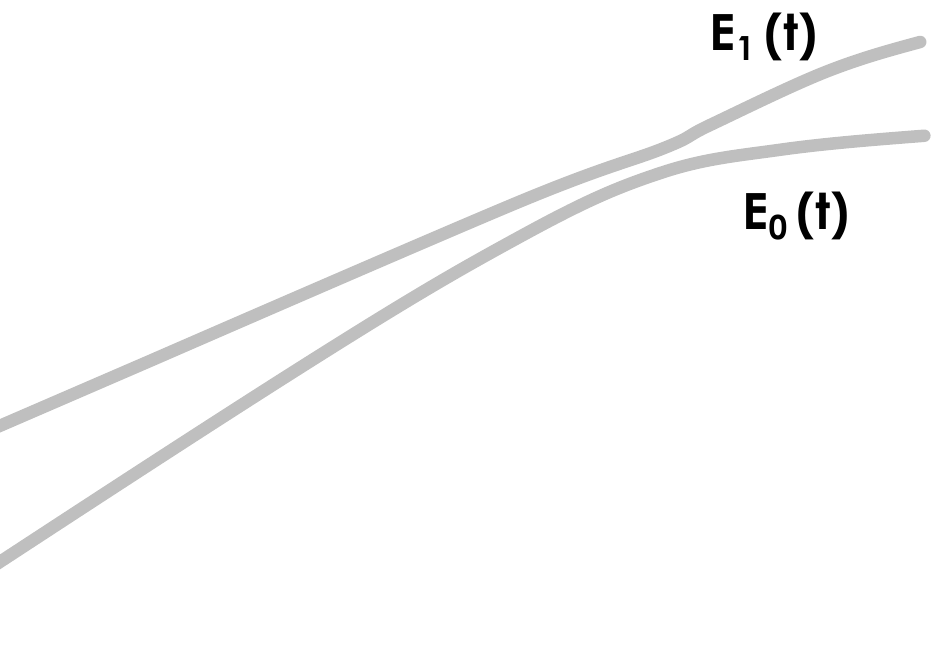}}&
 {\includegraphics[width=0.32\textwidth]{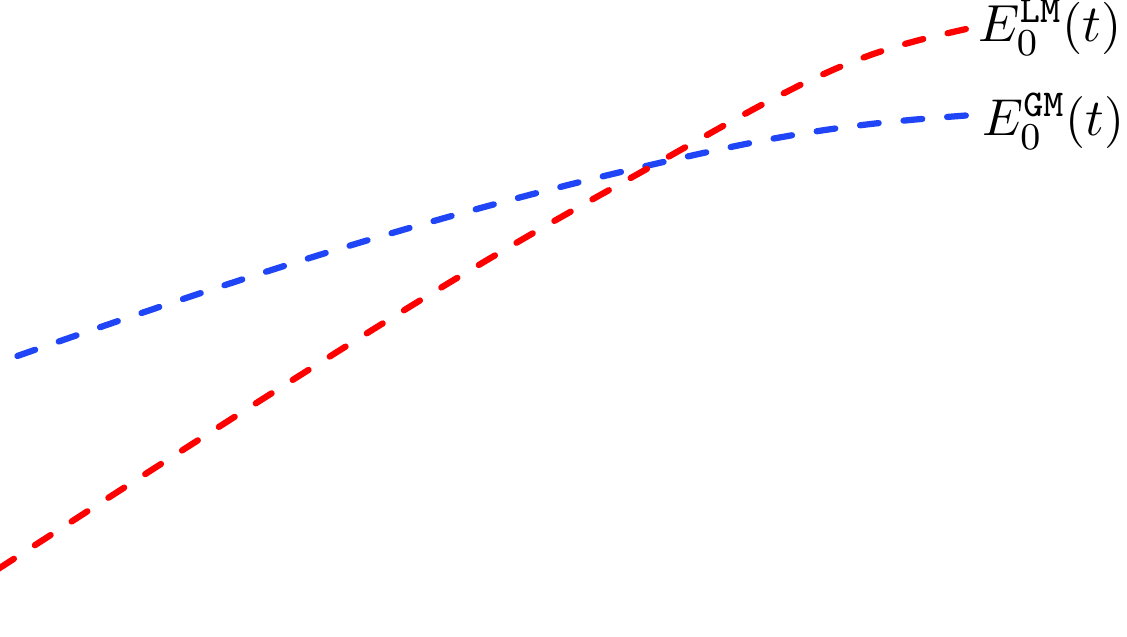}}&
 {\includegraphics[width=0.34\textwidth]{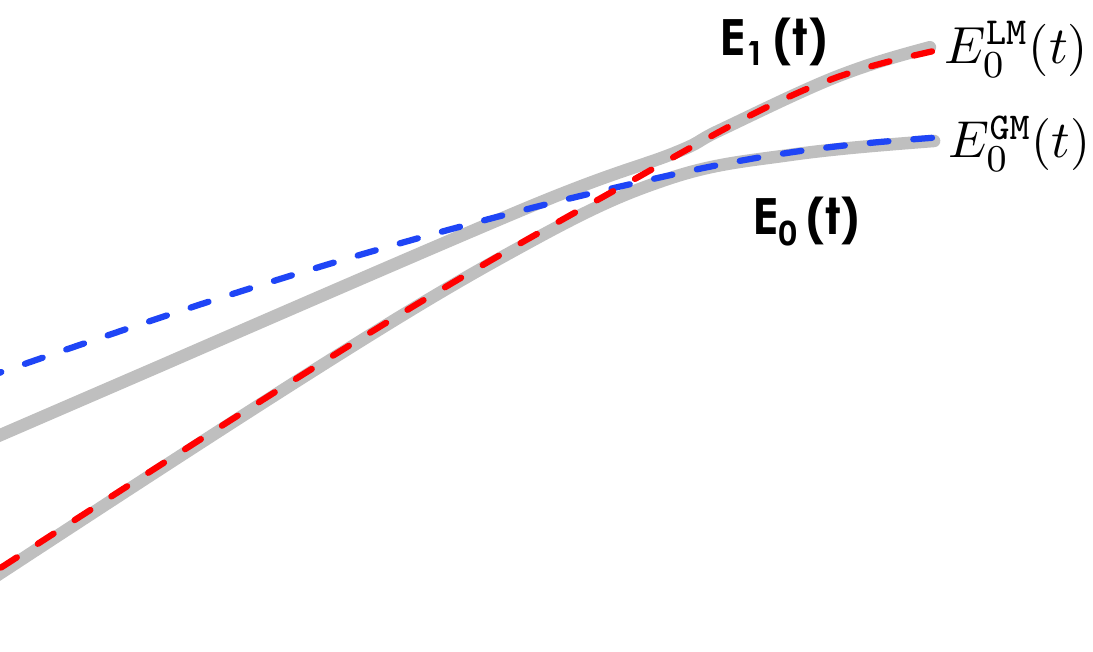}}\\
 (\textbf{a}) & (\textbf{b}) & (\textbf{c})
   \end{tabular}
   
  \begin{tabular}{ccc}
  {\includegraphics[width=0.32\textwidth]{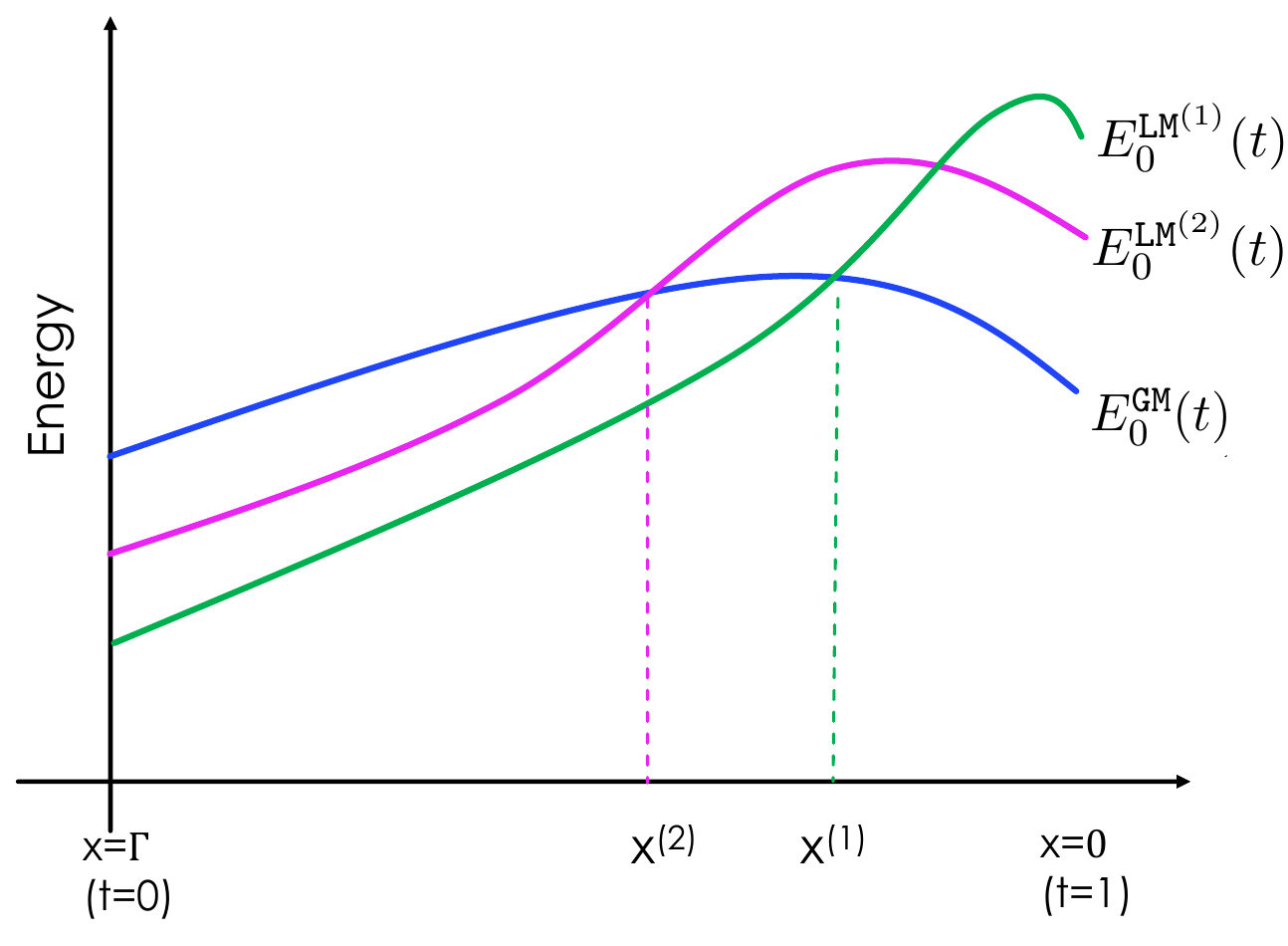}} &
  {\includegraphics[width=0.32\textwidth]{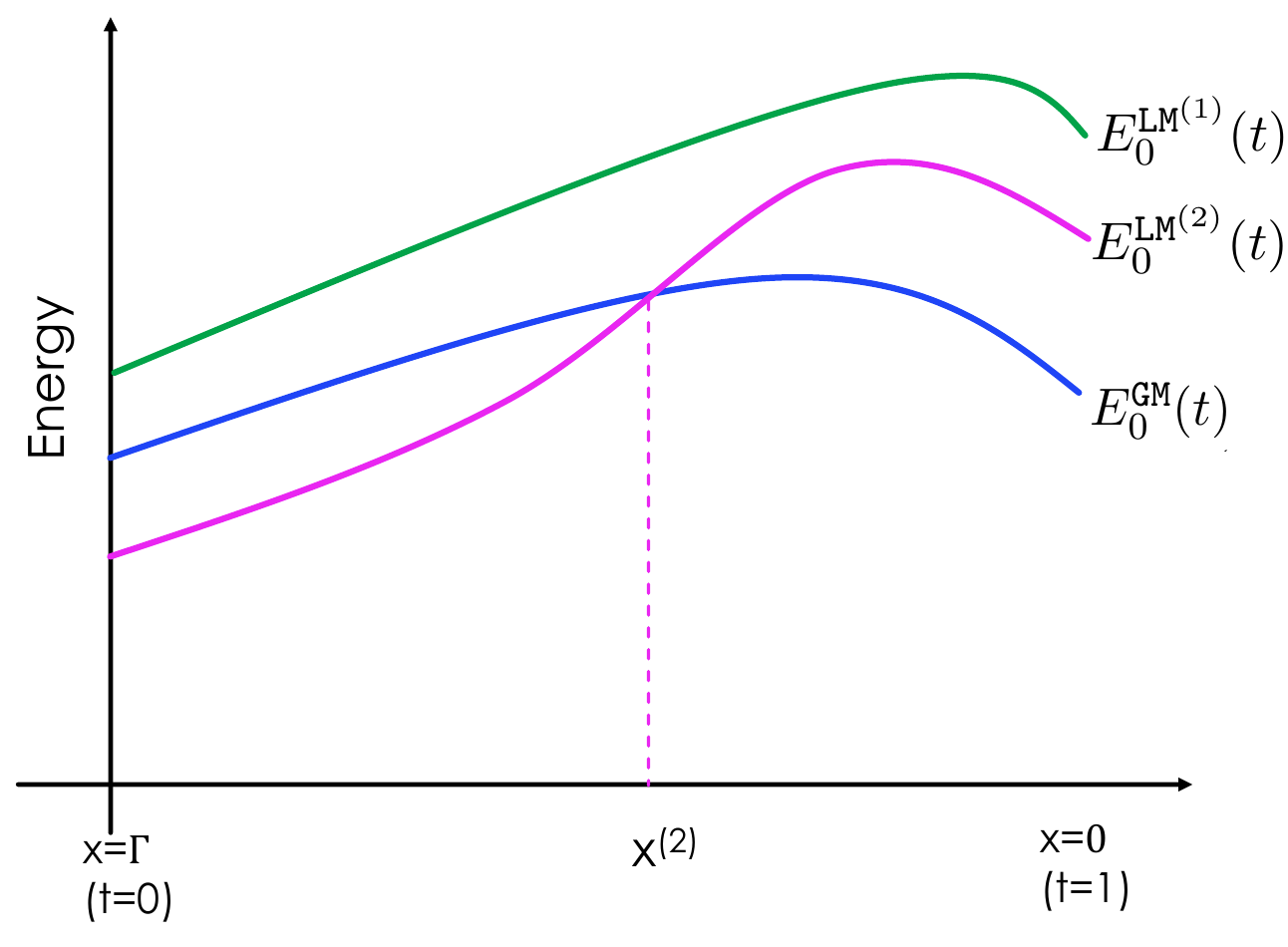}}&
  {\includegraphics[width=0.32\textwidth]{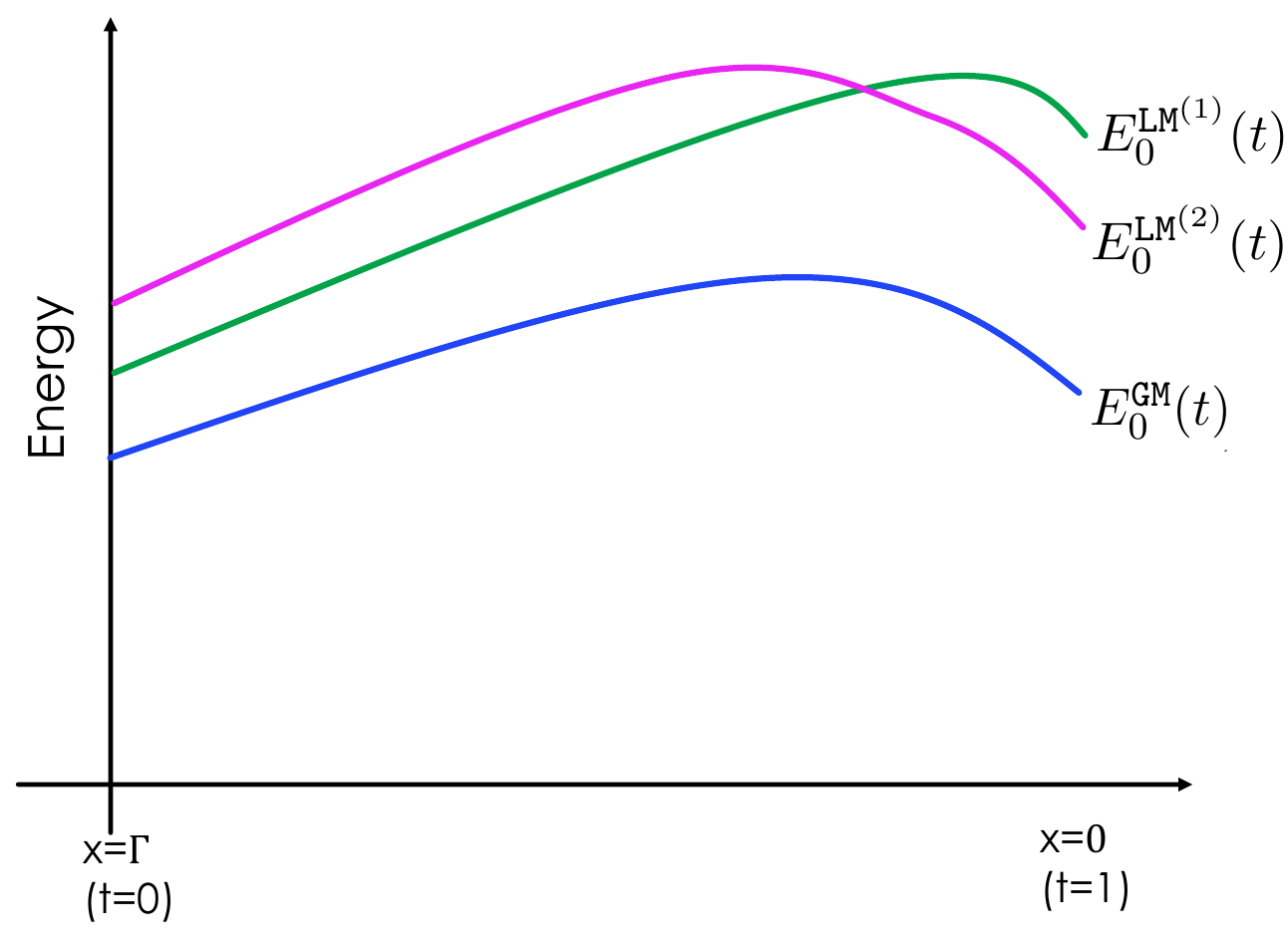}}\\
  (\textbf{d}) & (\textbf{e}) & (\textbf{f})
  \end{tabular}  

\caption{
\textbf{Anti-crossings and their iterative removal by \DDD{}.}
 \textbf{(a--c) Illustration of an $(\LM,\GM)$-anti-crossing.}
(\textbf{a}) An anti-crossing between the lowest two levels $E_0(t)$ and $E_1(t)$.  
(\textbf{b}) Bare energies $E_0^{\LM}(t)$ and $E_0^{\GM}(t)$ cross.  
(\textbf{c}) Overlay showing that the anti-crossing originates from this bare crossing.
 \textbf{(d--f) Conceptual illustration of the iterative removal of anti-crossings by \DDD{}.}
    (\textbf{d}) Under TFQA, the bare energy $E_0^{\GM}(t)$ (blue) crosses those of two local minima,
$E_0^{\LM^{(1)}}(t)$ and $E_0^{\LM^{(2)}}(t)$
    at $x^{(1)}=\mt{x}(t_1)$ and $x^{(2)}=\mt{x}(t_2)$.
    Each such bare crossing corresponds to an $(\LM^{(i)},\GM)$-anti-crossing in the system spectrum.
(\textbf{e}) After the first iteration of \DDD{}, the first crossing is removed by lifting
$E_0^{\LM^{(1)}}(t)$, while the second remains.
(\textbf{f}) A second iteration applies the same procedure to lift
$E_0^{\LM^{(2)}}(t)$, completing the removal of both anti-crossings.
}
 \label{fig:AC-iterative-removal}
\end{figure}

The \DDD{} algorithm was motivated by turning around the obstacle of stoquastic TFQA---specifically,
the presence of an anti-crossing caused by the competition between the
energies \emph{associated with} $\LM{}$ and $\GM$.
Informally, such an anti-crossing originates from a crossing of the bare
energies associated with $\LM$ and $\GM$, as illustrated in Figure~\ref{fig:AC-iterative-removal}(a--c).

The full algorithm consists of two phases; see Table~\ref{alg-overview}.
Phase~I uses polynomial-time TFQA to identify an independent-clique (IC)
structure associated with the critical $\LM$, which is responsible for the
(right-most) anti-crossing with $\GM$ in the TFQA ground state evolution.
Phase~II constructs an \( \XX \)-driver---based on the identified IC---to remove this anti-crossing.
This process is repeated iteratively: at each iteration, the algorithm identifies the next critical \LM{}, 
constructs an additional driver term, added to the system Hamiltonian.
In this way, the algorithm progressively removes small-gap obstructions from the system,
as illustrated conceptually in 
Figure~\ref{fig:AC-iterative-removal}(d--f).
Within this framework, the presence of an anti-crossing corresponds to exponentially slow adiabatic evolution,
whereas its removal leads to polynomial-time performance.

\begin{table*}[ht]
  \centering
  \caption{\textbf{Algorithm overview of the full \DDD{} algorithm.}}
\label{alg-overview}
\begin{tabular}{p{0.97\textwidth}}
\hline
\\[0em]
\textbf{Phase I: Independent-cliques extraction via polynomial-time TFQA.}
\begin{enumerate}
  \item Run TFQA with system Hamiltonian
  \( \mb{H}(t) = \x{t} \mb{H}_{\ms{X}} + \mt{p}(t)\mb{H}_{\ms{problem}} \)
  using a polynomial annealing time to return the excited states involved in the anti-crossing.
  \item Extract a set of seeds (local-minima states) from the TFQA output.
  \item Apply a classical post-processing procedure to identify an independent-cliques (IC) structure from these seeds.
\end{enumerate}

\textbf{Phase II: Non-stoquastic annealing with an $\XX$-driver.}
\begin{enumerate}
  \item Construct an $\XX$-driver graph \( G_{\ms{driver}} \) from the IC structure identified in Phase~I.
  \item Define a new time-dependent Hamiltonian
  \( \mb{H}(t) = \x{t} \mb{H}_{\ms{X}} + \jxx{t} \mb{H}_{\ms{XX}} + \mt{p}(t)\mb{H}_{\ms{problem}} \).
  \item For each feasible coupling strength \( \Jxx \), evolve the system using a three-stage schedule:
  \begin{itemize}
    \item \textbf{Stage~0:} initialization;
    \item \textbf{Stage~1:} energy-guided localization;
    \item \textbf{Stage~2:} interference-driven transition.
  \end{itemize}
  \item Measure the final state to extract the optimal solution.
\end{enumerate}
\\[0em]
\hline
\end{tabular}
\end{table*}

First, we recall the system Hamiltonian for \DDD{}:
\[
\mb{H}(t) = \x{t} \mb{H}_{\ms{X}} + \jxx{t} \mb{H}_{\ms{XX}} + \mt{p}(t)\mb{H}_{\ms{problem}},
\]
where
\(
\mb{H}_{\ms{X}} = - \sum_{i} \sigma_i^x, 
\mb{H}_{\ms{XX}} = \sum_{(i,j) \in \edge(G_{\ms{driver}})} \sigma_i^x \sigma_j^x,
\)
and 
$\mb{H}_{\ms{problem}}$ is the MIS-Ising Hamiltonian defined as
\begin{align}
\mb{H}_{\ms{problem}} 
&= \sum_{i \in \ver(G)} (-w_i) \shz{i} 
+ \Jzz^{\ms{clique}} \sum_{(i,j) \in \edge(G_{\ms{driver}})} \shz{i} \shz{j} 
+ \Jzz \sum_{(i,j) \in \edge(G) \setminus \edge(G_{\ms{driver}})} \shz{i} \shz{j}.
\label{eq:problem-Ham}
\end{align}
Here \( \shz{i} := \tfrac{I + \sigma^z_i}{2} \), where the coupling \( \Jzz^{\ms{clique}} \) is assigned to edges
corresponding to cliques in the driver graph, while \( \Jzz \) is assigned to all remaining edges.
The value of \( \Jzz^{\ms{clique}} \) is set sufficiently large to restrict the
system to the clique low-energy subspace, effectively cutting off the
high-energy sector, similar to the idea of the Projection Lemma
in~\cite{KKR2006}.
The time-dependent parameter schedule \( \jxx{t} \) depends on the \XX-coupling strength \( \Jxx \).
In particular, \( \jxx{t} \equiv 0 \) when \( \Jxx = 0 \), so this case reduces to TFQA without the \XX-driver.
The system Hamiltonian is stoquastic in the computational basis for \( \Jxx \le 0 \), and non-stoquastic for \( \Jxx > 0 \).

We now summarize how the resulting non-stoquastic evolution modifies the adiabatic path; a detailed analysis is given in Methods. The adiabatic evolution proceeds through three stages.
\begin{description}
\item 
  \textbf{Stage~0:} High-energy clique configurations are adiabatically suppressed, allowing restriction to a low-energy effective Hamiltonian.
\item \textbf{Stage~1:} The evolution is smoothly steered toward the \GM{}-support region without encountering an anti-crossing.
\item \textbf{Stage~2:} The original anti-crossing is removed through a see-saw spectral reshaping, enabling a smooth adiabatic path that can be explained through sign-generating quantum interference.
\end{description}
 %2405

\section{Discussion}

In this work, we identify a structurally defined family of MIS instances,
referred to as \GIC{} graphs, that are classically hard due to
the presence of extensive competing local minima organized by
independent-cliques substructures.
For these instances, we design and analyze a non-stoquastic adiabatic quantum
optimization algorithm that augments TFQA with a tailored $\XX$-driver.
The algorithm reshapes the low-energy spectrum
and removes the small-gap anti-crossings that arise in stoquastic evolution as
a consequence of tunneling, yielding a polynomial-time adiabatic path on
\GIC{} instances where both TFQA and state-of-the-art classical algorithms
exhibit exponential difficulty.
The essential quantum mechanism enabling the speedup is
the use of a non-stoquastic $\XX$-driver to access an enlarged,
sign-structured admissible subspace beyond the stoquastic regime.

More specifically, our analysis identifies two key mechanisms responsible for enabling the smooth
evolution path:
\begin{itemize}
\item \textbf{Structural steering}: energy-guided localization within the same-sign
block that steers the ground state smoothly into the \GM-supporting region,
bypassing tunneling-induced anti-crossings.

\item \textbf{Sign-generating quantum interference}: production of negative amplitudes
that enables an opposite-sign path through destructive interference in the
computational basis.
\end{itemize}
The analysis of these mechanisms is supported by both analytical derivations 
and numerical validation, and can, in principle, be made rigorous with 
further work.

\subsection*{Quantumness and Absence of Classical Analogues}
The emergence of negative amplitudes---produced by sign-generating interference due to the non-stoquastic Hamiltonian---serves as a witness to the quantum nature of the speedup.
Classical simulation algorithms, such as quantum Monte Carlo methods~(see \cite{Hen2021} and references therein), rely on the non-negativity of wavefunctions in the computational basis and break down in regimes where interference induces sign structure. This places the algorithm beyond the reach of (eventually) stoquastic annealing and efficient classical simulation.

From the perspective of classical solvers, the effect of sign-generating interference may be hard to reproduce classically.  
Algorithmically speaking, the \(\XX\)-driver with appropriately tuned coupling strength \(\Jxx\) enables a form of collective suppression of local minima, induced by edge couplings that effectively reduce vertex weights---while selectively preserving the global minimum, even in the presence of partial overlap.
To the best of our understanding, no known classical algorithm is able to replicate this effect, which appears to be a genuinely quantum capability.

Together, these points position \DDD{} as a candidate for demonstrating
quantum advantage, providing an explicit example of the emphasis expressed
in recent surveys of the field \cite{Huang2025}, where the origin of the
advantage can be traced to a specific quantum mechanism.

\subsection*{Relaxing the Structured Input Assumption}
Our analysis is developed under the structured \GIC{} assumption: each critical degenerate local minima---corresponding to a set of maximal independent sets of fixed size---is formed by a set of \emph{independent cliques} (\MIC{}). This assumption underlies both the design of the \(\XX\)-driver and the block decomposition of the Hamiltonian. The independence of cliques is assumed primarily to allow efficient identification during Phase~I.
In principle, one can allow the cliques to be dependent (\MDC{}), meaning that some edges are permitted between cliques. In this case, the cliques may need to be identified heuristically rather than exactly. Under suitable conditions, the algorithm may remain robust even when applied to \MDC{} structures.
More generally, each critical degenerate local minima in the real-world application
may consist of a set of
disjoint \MDC{}s. A clique-based driver graph may still be constructed in such cases, but the generalization becomes more intricate and may require further algorithmic refinements.
Other future directions include the study of weighted MIS instances and adaptive strategies for selecting or optimizing \(\Jxx\) during the anneal.

Our work thus aligns with the articulation that ``the most important reason to seek knowledge of instances with a quantum speedup is because an understanding of the type of structure that leads to advantage can help to find real-world problems where the advantage persists'' \cite{Babbush2025GrandChallenge}.
In this sense, the identification of structured MIS instances together with the associated quantum mechanism in \DDD{} offers a concrete instantiation of two central stages (Stage II and III) emphasized in \cite{Babbush2025GrandChallenge}.

From an algorithmic standpoint, AQC offers a
physics-driven framework for optimization.
By explicitly incorporating problem structure into the design of the system
Hamiltonian and by analyzing the resulting spectral behavior through the
presence or absence of anti-crossings, we demonstrate how AQC can be used as a
principled algorithm design paradigm rather than a black-box heuristic.
This approach leverages well-established tools from mathematical physics,
including perturbative analysis, effective Hamiltonian theory, and angular
momentum methods, in combination with algorithmic and graph-theoretic insights.

There are challenges in realizing the proposed non-stoquastic
algorithm directly on quantum annealing hardware, including constraints related
to hardware architecture, the minor-embedding problem~\cite{Choi2008,Choi2011},
and the requirements of error correction, as also noted
in~\cite{Barends2016NatureDAQC}.
Indeed, digitized AQC was promoted there as ``a viable quantum algorithm for
execution on an error-corrected digital quantum device.''
Direct Trotterization-based digitized implementations, however, may not be
resource-optimal.
Instead, one may adopt a QAOA-like approach~\cite{Farhi2014QAOA} to digitize the
\DDD{} algorithm, which can provide a more practical route forward.

Even prior to an immediate real-world application, our results are already valuable.
In particular, 
our analysis produces scalable small-scale models, derived
from our structural reduction, that capture the essential dynamics of
the algorithm.
These reduced models preserve the relevant block structure and interference
mechanisms while dramatically decreasing the effective system size,
with each large clique contracted to a single effective vertex.
As a result, the reduced instances are sufficiently small to be accessible on
current and near-term gate-model quantum processors through digitized AQC,
using Trotterized implementations.
This makes them well suited for validating hardware progress and for
experimental verification of the quantum advantage mechanism.
%975

\section{Methods}

The goal of the algorithm is to remove anti-crossings one by one using an
\XX-driver with appropriate choices of \( \Jxx \).
Detailed derivations, proofs, and numerical confirmations are provided in
Supplement.

We analyze how a single anti-crossing is encountered and removed.
As explained above, such an anti-crossing involves a critical \LM{} and the
global minimum (\GM{}).
The \LM{} corresponds to a degenerate set of maximal independent sets generated
by the cliques forming a \MIC{}, which are identified in Phase~I and used to
construct the \XX-driver.
We then restrict attention to reduced bipartite substructures formed by the
\LM{} and \GM{}.

\subsection{Phase I: Identification of the LM-associated clique structure}

In Phase~I, a polynomial-time transverse-field quantum annealing (TFQA)
procedure is used to identify the clique structure associated with a critical \LM{}.
Specifically, TFQA is run with the stoquastic Hamiltonian
\( \mb{H}(t) = \x{t}\mb{H}_{\ms{X}} + \mt{p}(t)\mb{H}_{\ms{problem}} \)
for a polynomial annealing time, and the resulting measurement outcomes are
recorded.

As shown in~\cite{Limit} (see Section~11 of Supplement for details),
an (\LM{},\GM{})-anti-crossing under TFQA exhibits an exponentially small gap.
As a consequence, polynomial-time TFQA undergoes a diabatic transition at the
anti-crossing and transitions into configurations associated with \LM{}.
The resulting output thus provides a collection of seed configurations
associated with \LM{}.

\begin{figure}[!htbp]
  \centering
  \begin{tabular}{cc}
     {\includegraphics[width=0.30\textwidth]{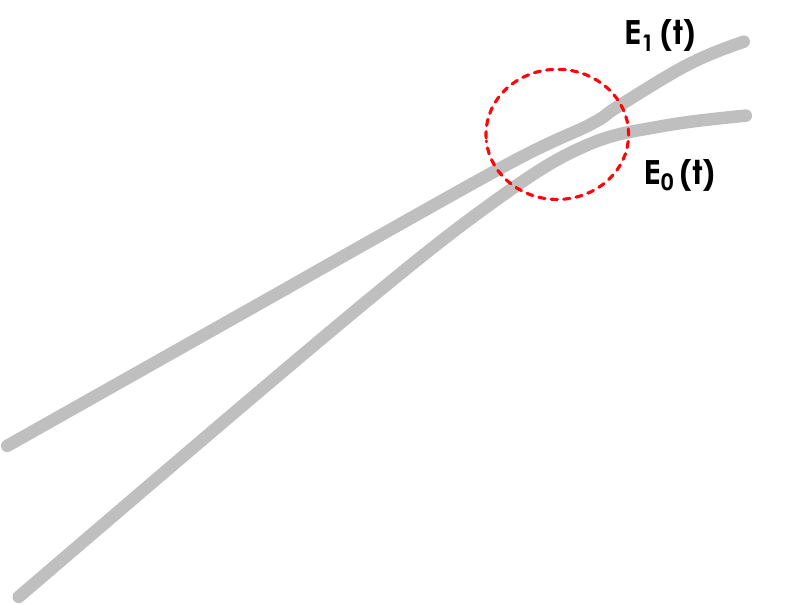}}&
     {\includegraphics[width=0.45\textwidth]{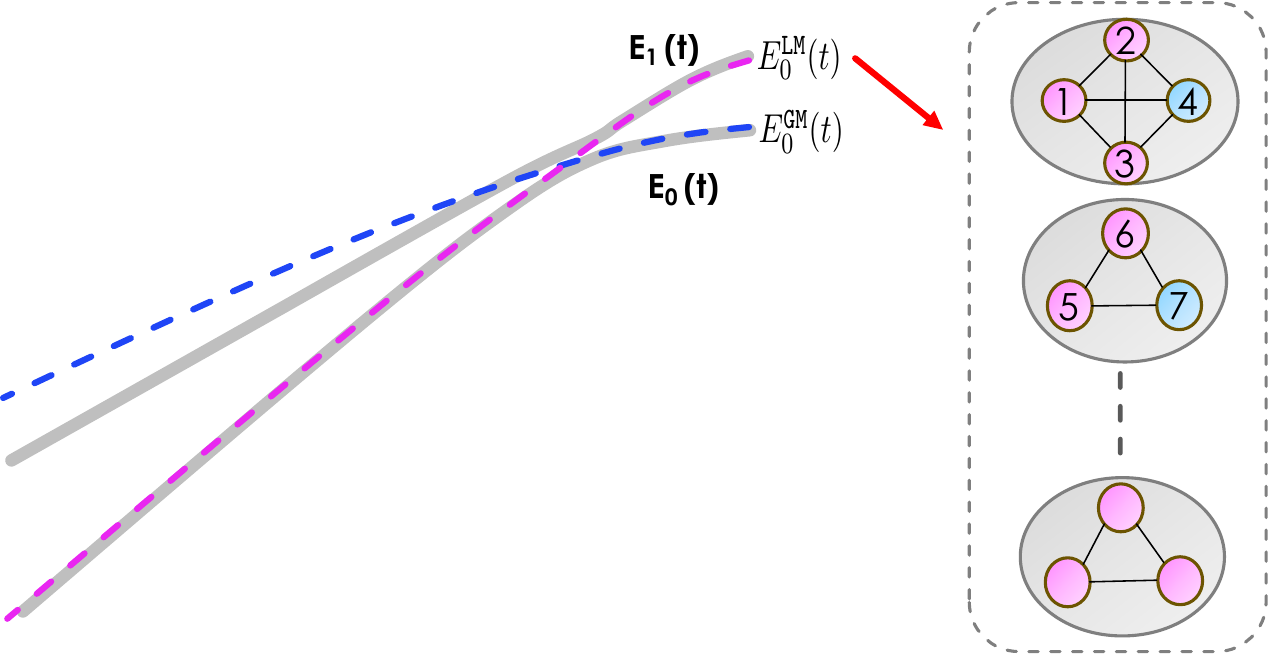}}\\
     (\textbf{a}) & (\textbf{b})
  \end{tabular}
    \caption{
\textbf{Schematic illustration of Phase~I.}
(\textbf{a}) An anti-crossing between the lowest two energy levels under stoquastic TFQA.
(\textbf{b}) The anti-crossing originates from competition between the energies associated with \LM{} and \GM{};
the \LM{} corresponds to a degenerate set of maximal independent sets generated
by cliques forming a \MIC{}.
}
 \label{fig:phase1}
\end{figure}

Under the structural assumption that the cliques forming the \MIC{} are mutually
independent, the underlying cliques structure can be identified efficiently by a
classical post-processing step.
The set of recovered cliques is then used to construct the \XX-driver graph for
Phase~II of the algorithm.
Thus, in what follows, \LM{}, \MIC{}, and the associated \XX-driver refer to the
same independent-cliques substructure.
The Phase~I mechanism is schematically illustrated in Fig.~\ref{fig:phase1}.

\subsection{Bipartite substructures: \Gdis{} and \Gshare{}}

To analyze the competition between a critical \LM{} and the \GM{}, we consider
two canonical bipartite substructures.
\begin{itemize}
  \item \textbf{\Gdis} (disjoint-structure), in which the vertex sets supporting
  the \LM{} and the \GM{} are disjoint.

  \item \textbf{\Gshare} (shared-structure), in which the \GM{} shares exactly
  one vertex with each clique comprising the \LM{}. 
\end{itemize}

We first define the \emph{disjoint-structure graph}
\( \Gdis = (V,E) \), in which the vertex set is partitioned into left and right
components with the following properties:
\begin{itemize}
  \item The left component consists of a set
  \( L = \{C_1,\dots,C_{m_l}\} \) of \( m_l \) disjoint cliques,
  each \( C_i = \cl(w_i,n_i) \), with vertex set
  \( V_L = \bigcup_{i=1}^{m_l} C_i \).
  \item The right component \( R \) consists of \( m_r \) independent vertices,
  each with weight \( w_r \).
  \item Every vertex in \( V_L \) is adjacent to every vertex in \( R \).
\end{itemize}

Throughout this work we focus on the unweighted MIS case and assume uniform
weights \( w_i = w_r = 1 \).
In the MIS--Ising energy landscape defined by $\mb{H}_{\ms{problem}}$,
the left component $V_L$ induces a degenerate set of local minima $\LM{}$
with degeneracy $\prod_{i=1}^{m_l} n_i$ and energy $-m_l$,
while the right component $R$ defines the global minimum $\GM{}$
of size $m_g = m_r$ and energy $-m_g$.

\begin{figure}[h]
  \centering
  \begin{tabular}{cc}
    \includegraphics[width=0.26\textwidth]{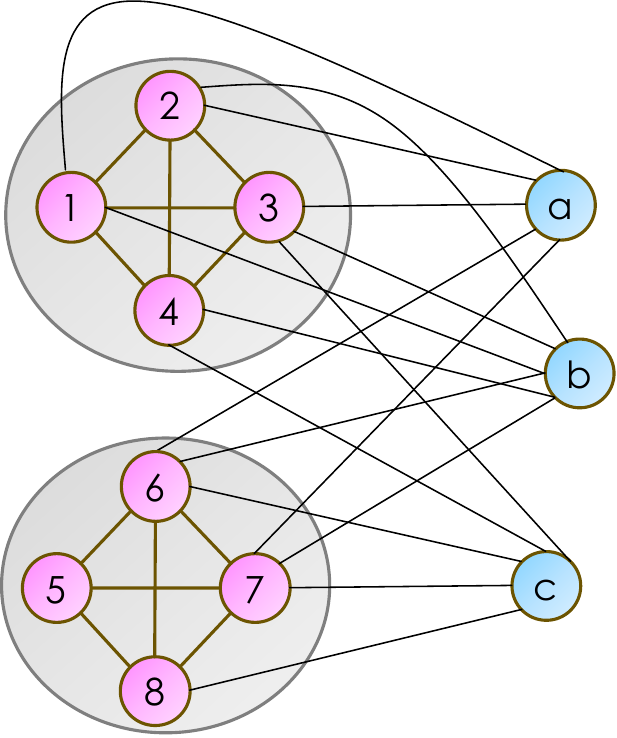}\quad \quad &
    \quad \quad \includegraphics[width=0.26\textwidth]{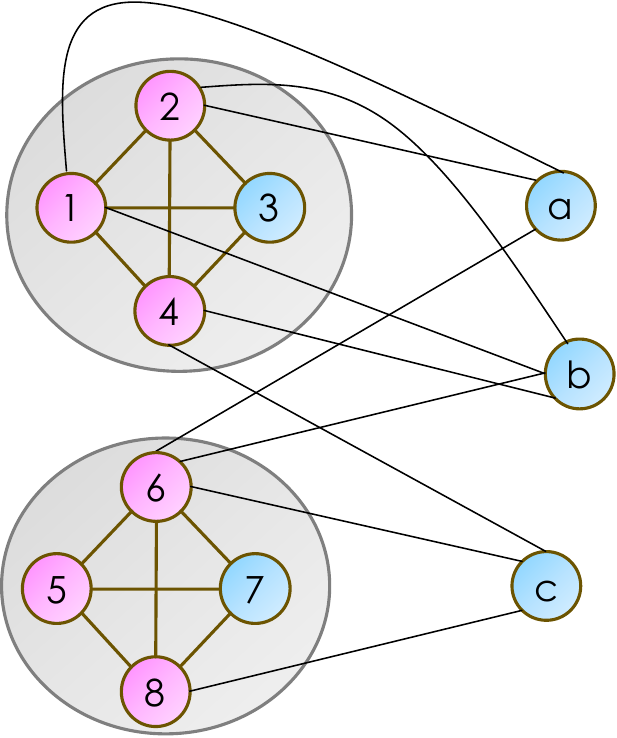}\\
    (\textbf{a}) & (\textbf{b})
  \end{tabular}
\caption{
\textbf{Example graphs illustrating the \Gdis{} and \Gshare{} structures.}
Recall that each \LM{} here is structurally a \MIC{}.
(\textbf{a}) Disjoint-structure graph \Gdis: The set \( L \) consists of \( m_l = 2 \) disjoint cliques,
each of size \( n_1 = n_2 = 4 \), whose vertices (pink) generate the local minima \LM{}.
The set \( R \) (blue) consists of \( m_r = 3 \) independent vertices forming the global minimum \GM{}.
(\textbf{a}) Shared-structure graph \Gshare: Relative to \Gdis, two vertices in \( L \) are shared between the \LM{} and \GM{},
and are therefore shown in blue instead of pink.
The global minimum (in blue) has \( m_g = 5 \) vertices.
In both cases, edges between the pink vertices in \( L \) and all vertices in \( R \) are complete,
though not all are shown for visual clarity.
}

\label{fig:dis-share-graph}
\end{figure}

The \emph{shared-structure graph} \( \Gshare \) differs from \( \Gdis \) in that
each vertex in \( R \) is adjacent to all but one vertex in each clique of \( L \).
This modification allows the \GM{} to include exactly one shared vertex from each
clique in \( L \), thereby introducing overlap between \LM{} and \GM{}.
The \GM{} then consists of these shared vertices together with all
vertices in \( R \), yielding total size \( m_g = m_l + m_r \).
The structures of \Gdis{} and \Gshare{} are illustrated in
Figure~\ref{fig:dis-share-graph}.

For convenience, we write \( m := m_l \) when no ambiguity arises.
We assume
\(
\sum_{i=1}^{m} \sqrt{n_i} > m_g,
\)
which ensures that the competition between \LM{} and \GM{} induces an
anti-crossing under stoquastic annealing.

We focus our Phase~II analysis on the shared-structure case \Gshare{},
  which captures the worst-case structure required for the full analysis.
  The disjoint-structure case \Gdis{} is included both to illustrate the underlying mechanism of \DDD{}---since
  the corresponding blocks of the effective Hamiltonian are decoupled and the analysis reduces to the same-sign block---and
  because it represents the worst-case structure for the analysis of anti-crossing gap under TFQA,
  as detailed in Section~11 of Supplement.

\subsection{Overview: Three-Stage Adiabatic Schedule in Phase II}

\begin{figure}[!htbp]
  \centering
  \includegraphics[width=0.65\textwidth]{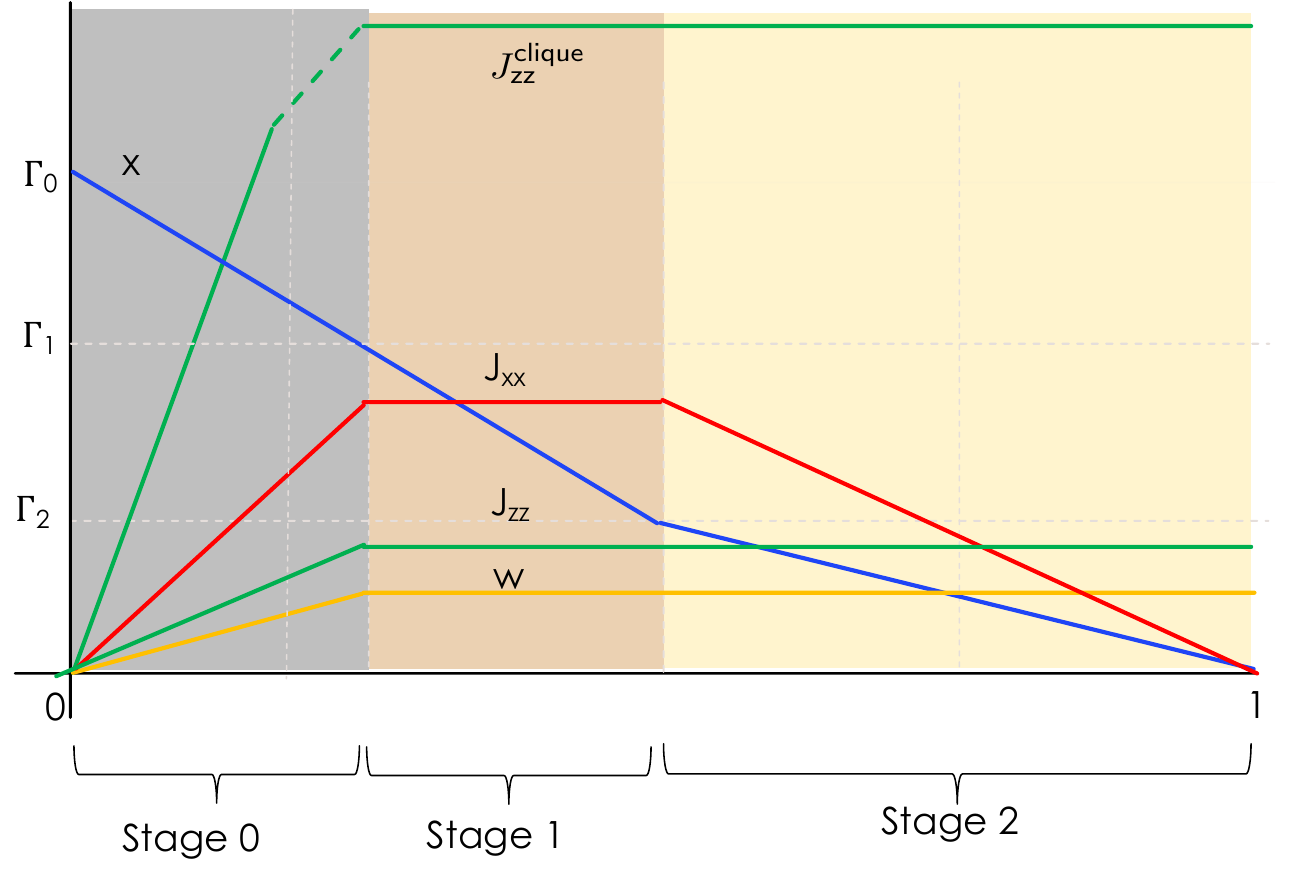}
  \caption{
  \textbf{Annealing parameter schedule for the system Hamiltonian  
  \( \mb{H}(t) = \x{t} \mb{H}_{\ms{X}} + \jxx{t} \mb{H}_{\ms{XX}} + \mt{p}(t)\mb{H}_{\ms{problem}} \).}
  The transverse field \( \x{t} \) (blue) begins at \( \Gamma_0 \),  
  decreases to \( \Gamma_1 \) in Stage~0, then to \( \Gamma_2 \) in Stage~1,  
  and reaches 0 at the end of Stage~2.
  The \( \XX \)-coupling strength \( \jxx{t} \) (red) increases linearly to \( \Jxx \) in Stage~0,  
  remains constant in Stage~1, and decreases to 0 in Stage~2.
  Vertex weights \( \mt{w}_i \) (orange) and $\ZZ$-couplings \( \Jzz, \Jzz^{\ms{clique}} \)  
  (green) ramp to their final values in Stage~0 and remain fixed thereafter.
  }
  \label{fig:schedule}
\end{figure}

In Phase~II, the \DDD{} Hamiltonian is decomposed into three stages, distinguished
by the relative ordering of the transverse-field strengths
\( \Gamma_0 > \Gamma_1 > \Gamma_2 \) and by when the problem and driver couplings
are activated, as illustrated by the annealing schedule in
Fig.~\ref{fig:schedule}.

This decomposition is conceptual and reflects the qualitative structure of the evolution; technical details are provided in the subsequent sections and in Supplement.

\paragraph{Stage~0 (Low-energy subspace formation).}
At early times, high-energy configurations containing intra-clique violations are adiabatically suppressed.
As a result, the dynamics become effectively restricted to the clique low-energy sector associated with the identified independent-clique structure.
This stage establishes an effective Hamiltonian that governs the subsequent evolution.

\paragraph{Stage~1 (Approach without anti-crossing).}
Within the resulting low-energy effective Hamiltonian, the system evolves toward the \GM{}-support region while avoiding localization regimes associated with local minima. In this stage, the evolution does not encounter an anti-crossing, ensuring that the system reaches the parameter regime where the non-stoquastic effects become operative.

\paragraph{Stage~2 (Anti-crossing removal).}
In the final stage, the remaining low-energy competition between \LM{} and \GM{} is resolved. The \XX-coupling induces a see-saw reshaping of the spectrum, lifting same-sign local minima while allowing opposite-sign sectors to participate in the evolution. The resulting adiabatic path remains smooth and can be understood in terms of sign-generating quantum interference.

We now analyze each stage in turn, beginning with the construction of the effective low-energy Hamiltonian in Stage~0.

\subsection{Stage~0: Effective Low-Energy Hamiltonian}

During Stage~0, a large energy penalty 
\( \Jzz^{\ms{clique}} \) is applied to the intra-clique edges, where the cliques
are those that constitute the \XX-driver graph.
This separates the Hilbert space into a low-energy sector, consisting of
configurations that respect the clique constraints,
and a high-energy sector containing states that incur intra-clique energy penalties.

With respect to this decomposition, the system Hamiltonian admits a block form,
\[
  \mb{H} \;=\;
  \begin{pmatrix}
    \mb{H}^{\ms{low}} & \mb{V} \\
    \mb{V}^\dagger & \mb{H}^{\ms{high}}
s  \end{pmatrix},
\]
where \( \mb{H}^{\ms{low}} \) and \( \mb{H}^{\ms{high}} \) denote the projections
onto the low-energy and high-energy sectors, respectively.

As a result, the evolution at the end of Stage~0 is governed by an effective
low-energy Hamiltonian
$\Heff = \mb{H}^{\ms{low}}$.
A detailed derivation of this effective Hamiltonian, together with a proof that
the spectral gap remains large throughout Stage~0, is provided in Section~8 of
Supplement.
From this point onward, all block decompositions and dynamical analyses refer to
the time-dependent effective Hamiltonian \( \Heff \) governing Stages~1 and~2.

\subsection{Angular-momentum block decomposition and $\Jxx$-induced energy reshaping}

We develop a block decomposition of the effective Hamiltonian into same-sign and
opposite-sign sectors in the angular-momentum basis induced by the cliques in
the \XX-driver graph.

Within this basis, the \XX-coupling term acts diagonally.
In particular, the coupling strength $\Jxx$ shifts the energies of same-sign and
opposite-sign blocks with opposite signs.
This reshapes the spectrum in a see-saw fashion: increasing $\Jxx$ raises the
energies associated with $\LM{}$ in the same-sign block while lowering those of
the opposite-sign blocks.
With an appropriate choice of $\Jxx$, this energy reshaping removes the
$(\LM{},\GM{})$-anti-crossing under stoquastic evolution.

In the following, we first analyze the angular-momentum structure and $\Jxx$-induced energy
reshaping for a single clique, and then extend the analysis to a collection of
independent cliques (a \MIC{} bare subsystem) and finally to the full effective Hamiltonian.

\subsubsection{Single-clique analysis}
\label{sec:single-clique}
Let \( \SLE \) denote the low-energy subspace of a single clique, \( \ms{Clique}(w_c, n_c) \),  
where \( w_c \equiv 1 \) denotes the vertex weight and \( n_c \) the clique size.
The low-energy subspace \( \SLE \), which consists of the \( n_c + 1 \) independent-set
configurations of the clique, admits a natural decomposition in the total
angular-momentum basis into a single effective spin-\(\tfrac{1}{2}\)
(same-sign) sector and a collection of spin-0 (opposite-sign) sectors:
\begin{align}
  \SLE
  =
  \left[\tfrac{1}{2}\right]_{n_c}
  \oplus
  \underbrace{0 \oplus \cdots \oplus 0}_{n_c - 1}.
  \label{eq:single-clique}
\end{align}
The spin-\(\tfrac{1}{2}\) sector consists of two same-sign basis states, while
each spin-0 sector consists of a single opposite-sign basis state.

\begin{figure}[!htbp]
  \centering
 \begin{tabular}{cc}
 \includegraphics[width=0.4\linewidth]{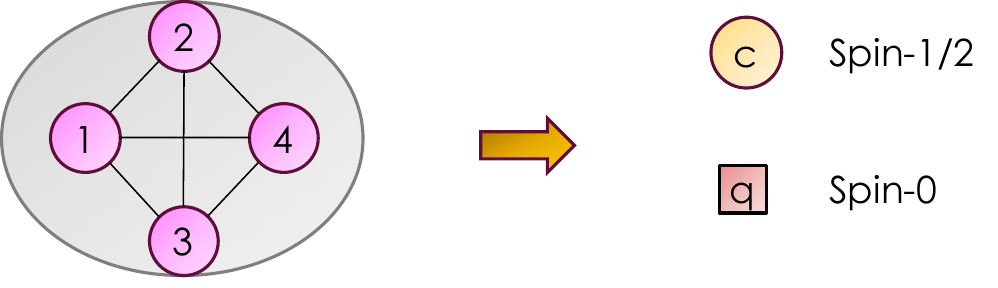} &
 \includegraphics[width=0.5\linewidth]{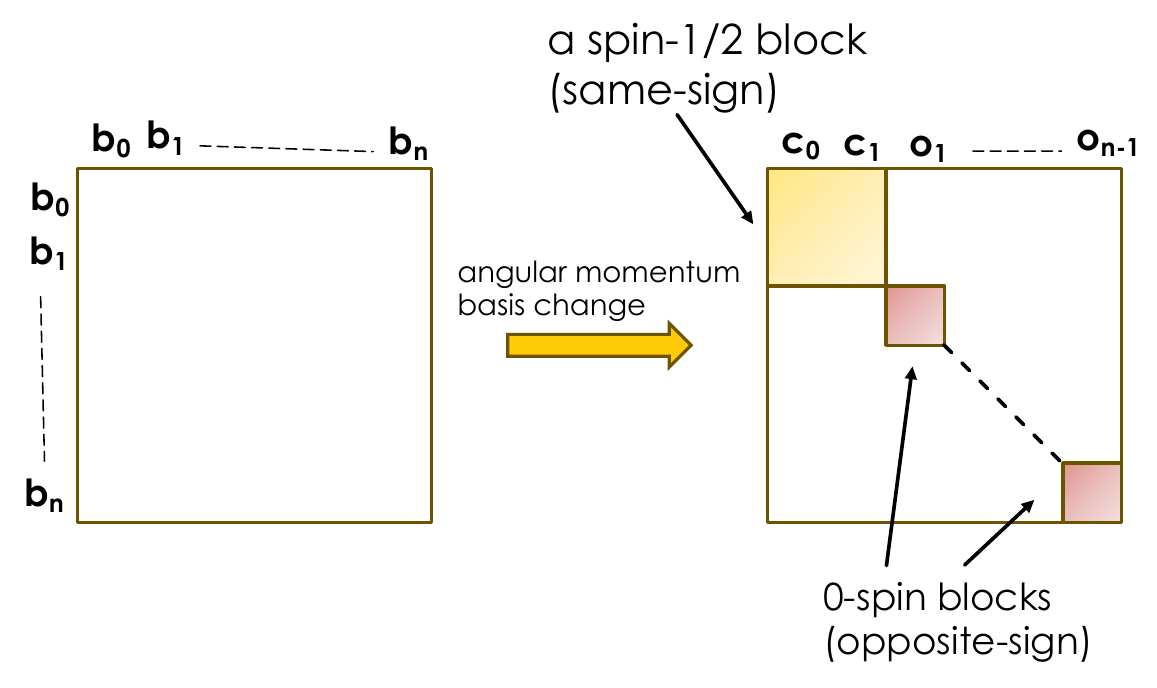}\\
 (\textbf{a}) & (\textbf{b})
  \end{tabular}
  \begin{tabular}{cc}
    \includegraphics[width=0.5\linewidth]{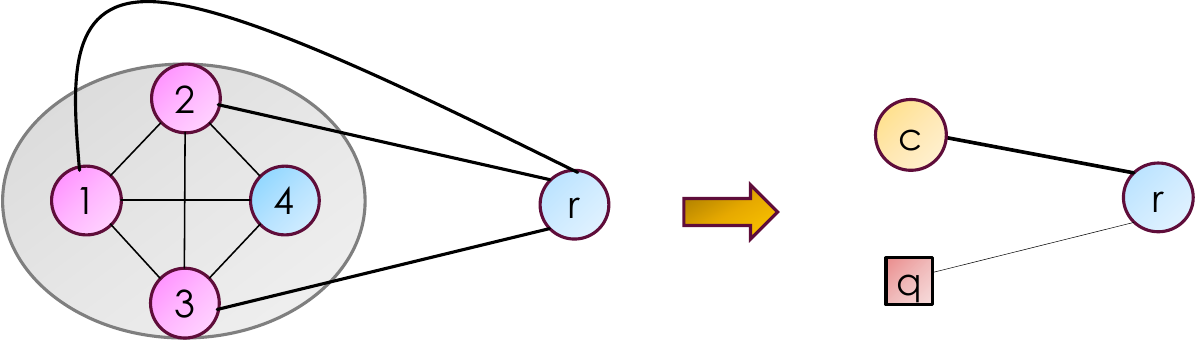}&
    \includegraphics[width=0.4\linewidth]{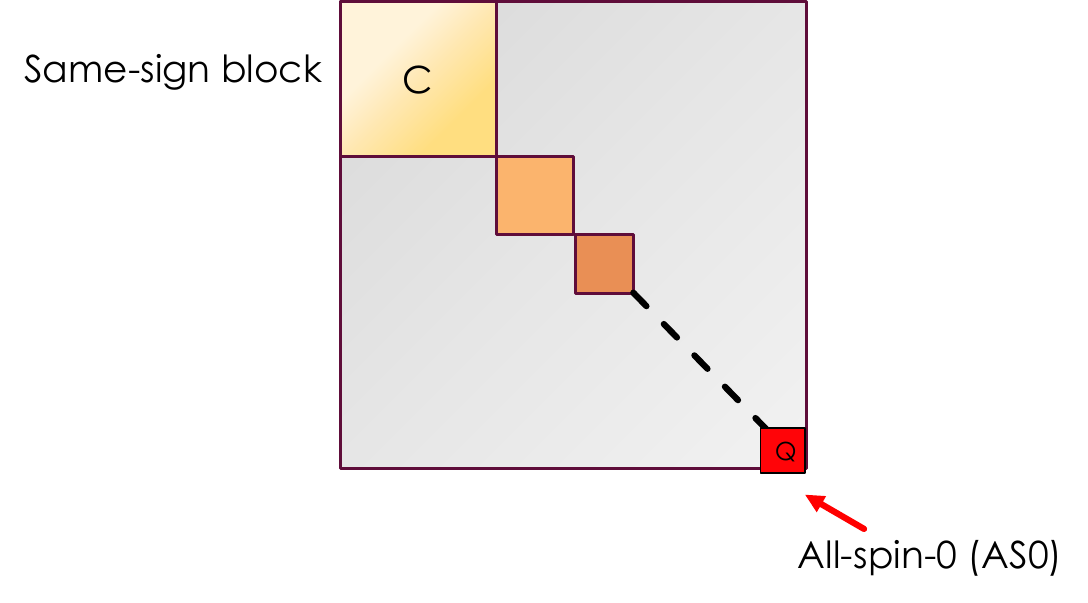} \\
    (\textbf{c}) & (\textbf{d})
  \end{tabular}
\caption{\textbf{Block decomposition in angular momemtum basis.}
  (\textbf{a}) A clique (\(\{1,2,3,4\}\)) is transformed into an effective spin-\( \tfrac{1}{2} \) component labeled \(c \) (orange)  
  and a representative spin-0 component labeled \( q \) (red square).
 (\textbf{b}) \textbf{Single clique.}The low-energy product basis decomposes into an effective spin-\(\tfrac{1}{2}\)
  same-sign block and \(n-1\) spin-0 opposite-sign blocks in the angular-momentum basis.
(\textbf{c}) Partial adjacency of an external vertex \( r \) to a clique leads to
 \(\ZZ\)-couplings to both the same-sign and
opposite-sign sectors, inducing mixing between them.
(\textbf{d}) \textbf{Full system.} Tensoring the single-clique structure yields a block decomposition of the effective
  Hamiltonian \( \bar{\Heff} \) into a dominant same-sign block
  \( \mb{H}_{\mc{C}} \) and opposite-sign blocks, including the AS0 block
  \( \mb{H}_{\mc{Q}} \); these blocks are decoupled in \Gdis{} and coupled via
  \( \Hinter \) in \Gshare{}.
  }
  \label{fig:block-decomp}
\end{figure}

An illustration of this basis transformation is shown in
Figure~\ref{fig:block-decomp}(a,b).

The effective Hamiltonian on the spin-\( \tfrac{1}{2} \) (same-sign) sector is a
\(2\times2\) matrix of the form
\[
\mb{B}\!\left( \mt{\weff_c},\, \sqrt{n_c}\,\mt{x} \right)
=
\begin{bmatrix}
 -\!\left( w_c - \tfrac{n_c - 1}{4}\, \mt{jxx} \right)
 &
 -\tfrac{\sqrt{n_c}}{2}\, \mt{x}
 \\[6pt]
 -\tfrac{\sqrt{n_c}}{2}\, \mt{x}
 &
 0
\end{bmatrix}
\]
with \(
\mt{\weff_c}
=
w_c - \tfrac{n_c - 1}{4}\mt{jxx},
\)
while each spin-0 (opposite-sign) sector contributes a single eigenvalue
\(
- (w_c + \tfrac{1}{4}\, \mt{jxx}).
\)

\begin{figure}[!htbp]
\centering
\begin{tabular}{cc}
\includegraphics[width=0.4\textwidth]{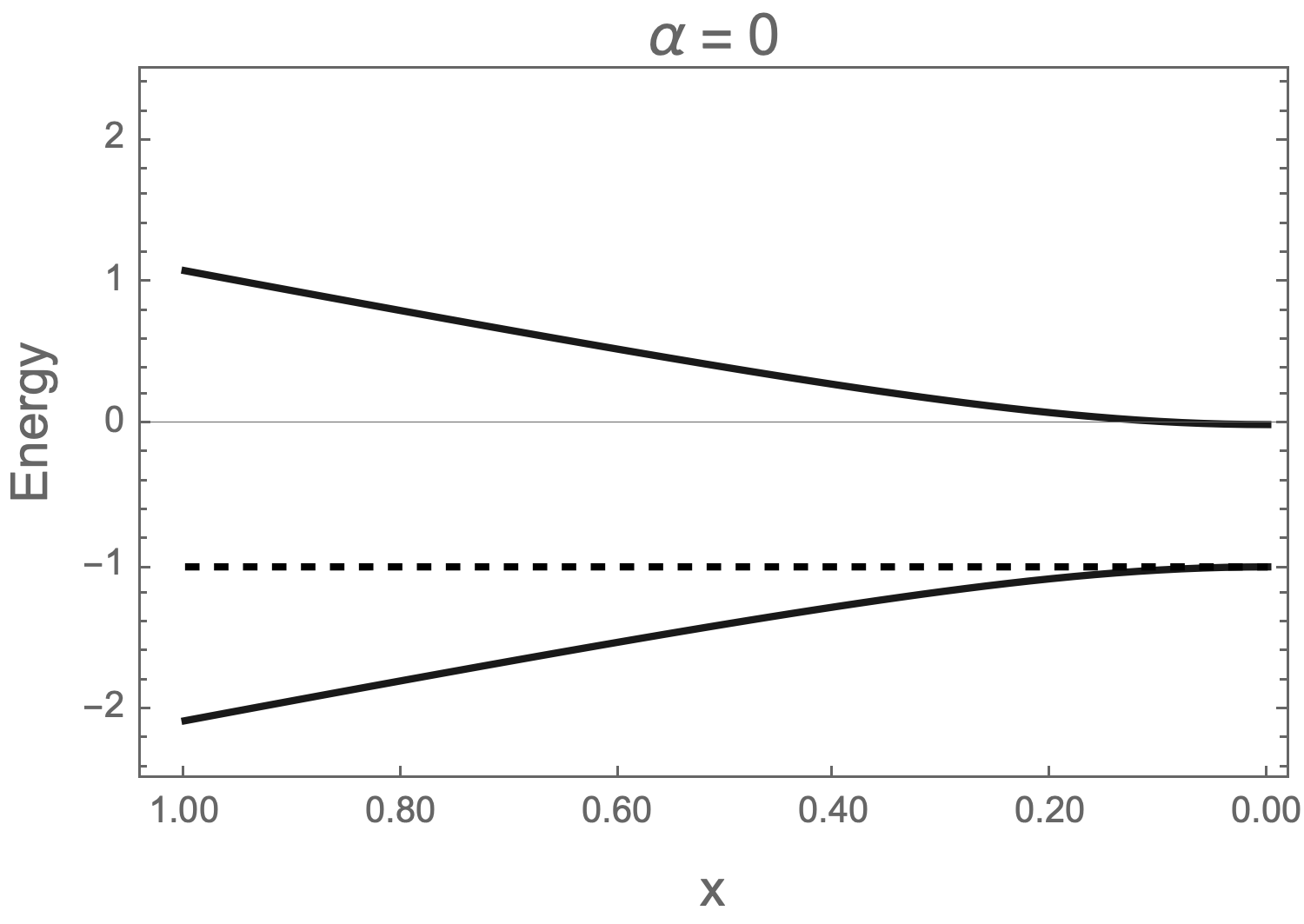} &
\includegraphics[width=0.4\textwidth]{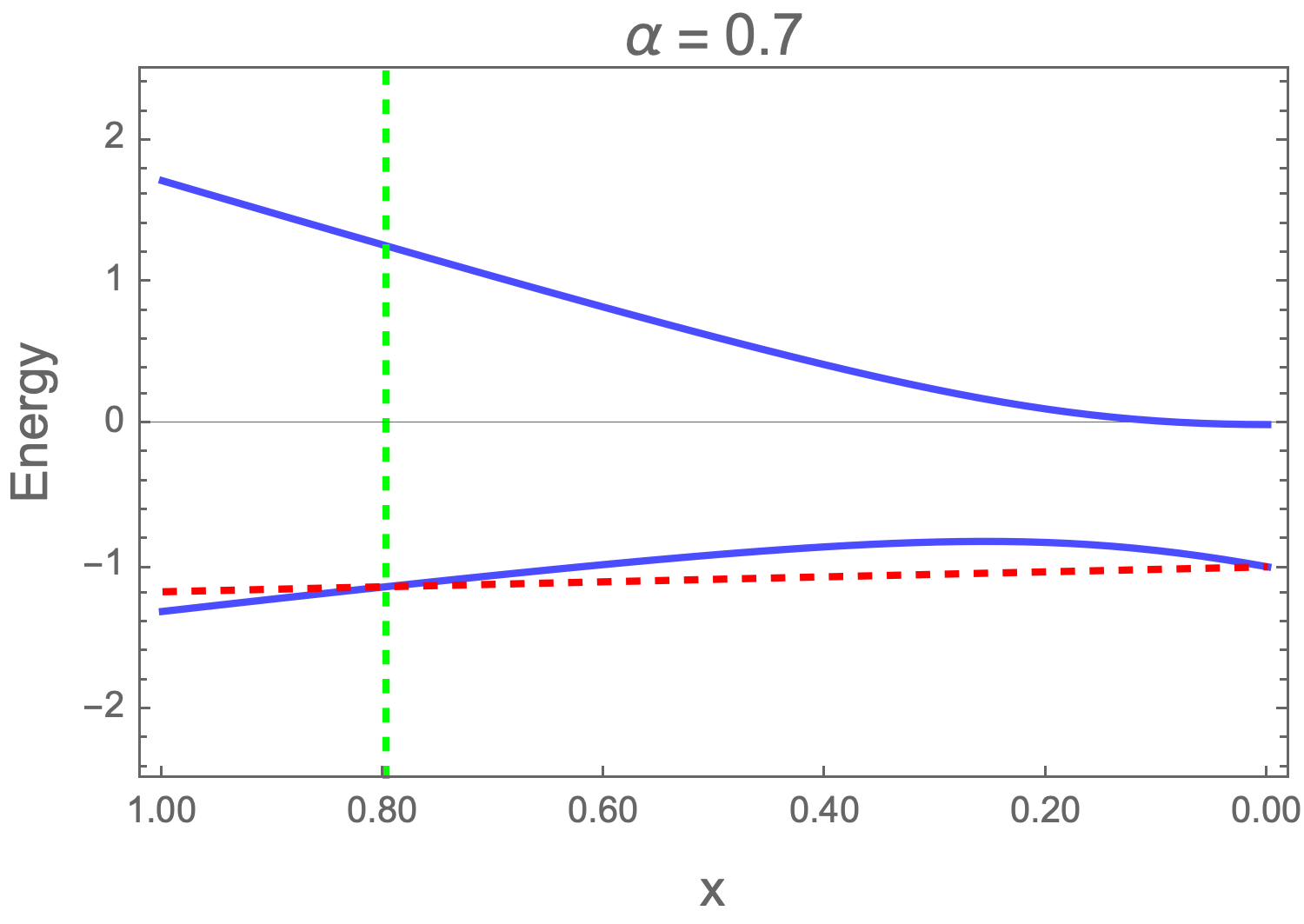} \\
(\textbf{a}) & (\textbf{b})\\
\end{tabular}
\includegraphics[width=0.8\linewidth]{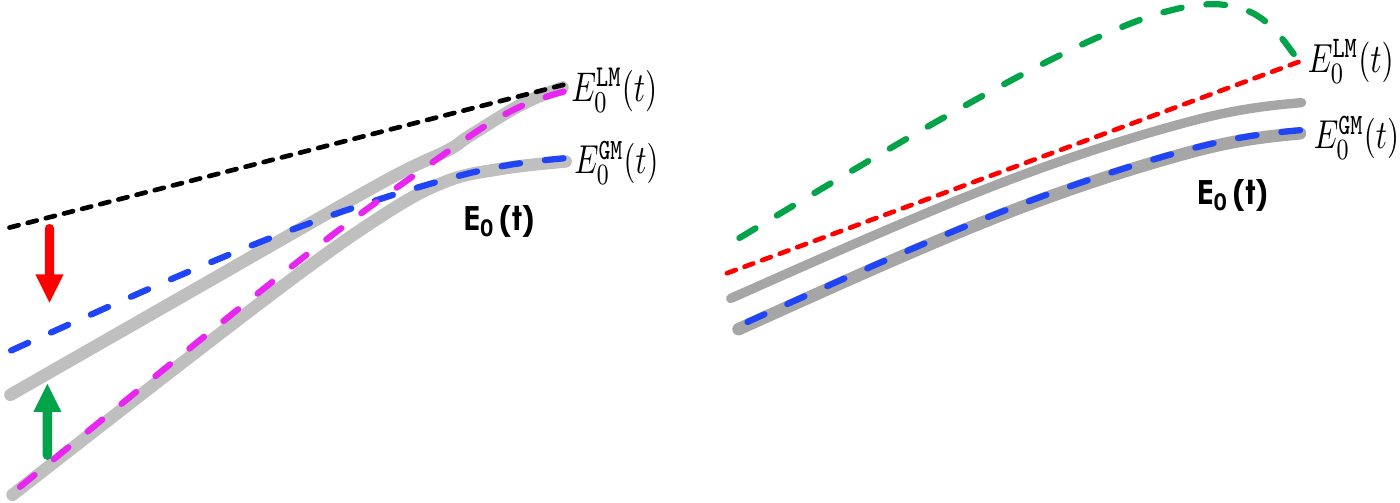}\\
(\textbf{c})\\
 \caption{\textbf{See-saw reshaping of the spectrum induced by $\Jxx$.}
(\textbf{a--b}) Effect of the \XX-coupling strength $\Jxx$ on the low-energy spectrum of a
single clique, with $\Jxx = \alpha \Gamma_2$.
The transverse field $\mt{x}$ decreases from $\Gamma_2 = 1$ to $0$, with
$\mt{jxx} = \alpha \mt{x}$.
\textbf{See-saw effect:} increasing $\Jxx$ raises the energies of the same-sign
sector (black to blue) while lowering the energy of the opposite-sign sector
(black dashed to red dashed).
(\textbf{c}) Schematic illustration of the resulting see-saw mechanism in the full
system: for $\Jxx = 0$, an $(\LM{},\GM{})$-anti-crossing appears under stoquastic
TFQA, while for a feasible nonzero $\Jxx$, the same-sign $\LM$ energy is lifted (pink dashed to green dashed) and
the lowest opposite-sign energy is lowered (black dotted to red dotted), removing the anti-crossing.
}
\label{fig:see-saw}
\end{figure}

Notice that the \XX-coupling $\mt{jxx}$ is diagonal and acts with
opposite sign on the two sectors.
It reshapes the spectrum in a see-saw fashion:
 the same-sign sector is lifted in energy (and its slope magnitude is reduced), while the
opposite-sign sectors are lowered. See Figure~\ref{fig:see-saw}(a).

Moreover, this angular-momentum decomposition can be expressed in terms of
coupled subcliques, as illustrated in Figure~\ref{fig:block-decomp}(c),
which allows treating non-uniform \(\ZZ\)-couplings to
external vertices and is essential for the global block decomposition of graphs
with shared structure.

The details are provided in Section~6 of Supplement.

\subsubsection{Bare subsystem analysis}
\label{sec:bare-sub}

We now extend the single-clique analysis to a collection of independent cliques (a \MIC{}), which
together forms a \emph{bare subsystem}.
It is a \emph{subsystem} because we consider only the subgraph induced by the \MIC{},  
and it is \emph{bare} because it is fully decoupled from the rest of the graph.
Note that either \LM{} or \GM{} by itself forms a bare subsystem, where each clique in \GM{} has size one.

Consider the bare subsystem formed by \( m \) independent cliques \( \ms{Clique}(w_i, n_i) \),  
where \( w_i \equiv 1 \) denotes the vertex weight and \( n_i \) the size of the \( i \)th clique.  
%This graph admits \( \prod_{i=1}^{m} n_i \) degenerate maximal independent sets of size \( m \).

The low-energy space of the bare subsystem decomposes into three classes of
sectors: the \emph{same-sign sector} \( \mc{C}^{\ms{bare}} \), the
\emph{intermediate opposite-sign sectors} \( \mc{W}^{\ms{bare}} \), and the
\emph{all-spin-zero (AS0) opposite-sign sector} \( \mc{Q}^{\ms{bare}} \).

%\paragraph{Same-sign and opposite-sign sectors.}

From the single-clique result (Eq.~\eqref{eq:single-clique}), the low-energy subspace of each clique decomposes as:
\[
\mc{L}_i = \left[\tfrac{1}{2}\right]_{n_i} \oplus \underbrace{0 \oplus \cdots \oplus 0}_{n_i - 1}.
\]

Hence, the total low-energy subspace \( \mc{L} \) of the bare subsystem is given by:
\[
\mc{L} = \bigotimes_{i=1}^{m} \mc{L}_i 
= \bigotimes_{i=1}^{m} \left( \left[\tfrac{1}{2}\right]_{n_i} \oplus \underbrace{0 \oplus \cdots \oplus 0}_{n_i - 1} \right)
= \bigoplus \left( \bigotimes_{i=1}^{m} \left[\tfrac{1}{2} \text{ or } 0 \right] \right),
\]
where the direct sum ranges over all tensor products selecting one spin-\( \tfrac{1}{2} \) and \( n_i - 1 \) spin-zeros from each clique.

This decomposition yields \( \prod_{i=1}^{m} n_i \) block subspaces. Among them, the block
\[
\mc{C}^{\ms{bare}} := \bigotimes_{i=1}^{m} \left[\tfrac{1}{2}\right]_{n_i}
\]
is composed entirely of spin-\( \tfrac{1}{2} \) factors and is referred to as the \emph{same-sign sector}.

At the opposite extreme, there are \( \prod_{i=1}^{m} (n_i - 1) \) blocks composed entirely of spin-$0$ factors:
\[
\mc{Q}^{\ms{bare}} := \bigotimes_{i=1}^{m} 0,
\]
referred to as the \emph{all-spin-zero (AS0) opposite-sign sector}.

The remaining opposite-sign blocks, each containing a mixture of spin-\( \tfrac{1}{2} \) and spin-$0$ components, are called the \emph{intermediate sectors}.
Let \( \mc{W}^{\ms{bare}} \) denote one such intermediate sector.

%\paragraph{Bare-subsystem Hamiltonians.}

With this sector decomposition, projecting the effective Hamiltonian
yields a block-diagonal structure.

In particular, the restriction to the same-sign sector takes the form
%\begin{equation}
  \[
\mb{H}_{\mc{C}^{\ms{bare}}}
=
\sum_{i=1}^{m}
\mb{B}_i\!\left( \mt{\weff_i},\, \sqrt{n_i}\,\mt{x} \right),
\]
%\label{eq:bare-samesign}
%\end{equation}
where each \( \mb{B}_i \) acts nontrivially only on the \(i\)th clique and
\(
\mt{\weff_i}
=
w_i - \tfrac{n_i - 1}{4}\mt{jxx}.
\)

The AS0 opposite-sign sector contributes a single energy level,
\[
\mb{H}_{\mc{Q}^{\ms{bare}}}
=
-\sum_{i=1}^{m}
\left(
w_i + \tfrac{1}{4}\mt{jxx}
\right),
\]
while each intermediate opposite-sign sector corresponds to a reduced same-sign
block involving fewer active \(\mb{B}_i\) terms, together with an additive energy
shift.

Therefore, the bare subsystem has a collective \( \Jxx \)-induced see-saw reshaping
of its spectrum, directly as a sum of single-clique contributions. Furthermore,
the closed-form solutions of the energies (in particular, $E_0^{\LM}(t)$ and $E_0^{\GM}(t)$)  enable the derivation of analytic bounds on \( \Jxx \) required to remove the
\((\LM,\GM)\)-anti-crossing.

The details are provided in Section~7 of Supplement.

\subsubsection{Block decomposition of the effective Hamiltonian}

Finally, we assemble the full effective Hamiltonian by combining the $L$ bare-subsystem structure
with the remaining degrees of freedom in $R$.
The same-sign sector of the full system is defined as
\[
\mc{C}
=
\mc{C}^{\ms{bare}} \otimes \left( \mathbb{C}^2 \right)^{\otimes m_r},
\]
where \( m_r = |R| \).
Similarly, the opposite-sign sectors of the full system are given by
\[
\mc{W}
=
\mc{W}^{\ms{bare}} \otimes \left( \mathbb{C}^2 \right)^{\otimes m_r},
\qquad
\mc{Q}
=
\mc{Q}^{\ms{bare}} \otimes \left( \mathbb{C}^2 \right)^{\otimes m_r}.
\]
Accordingly, \( \mc{Q} \) is also referred to as the all-spin-zero (AS0)
opposite-sign sector of the full system.

With these definitions, the effective Hamiltonian decomposes into a same-sign
block \( \mb{H}_{\mc{C}} \) and multiple opposite-sign blocks
\( \mb{H}_{\mc{W}} \) and \( \mb{H}_{\mc{Q}} \), which are either decoupled (in the
disjoint-structure case) or coupled (in the shared-structure case), as illustrated schematically in
Figure~\ref{fig:block-decomp}(d).

The explicit block Hamiltonians, coefficients, and inter-block couplings are
derived in Section~9.1 of Supplement.

The initial ground state resides in the same-sign block \( \mb{H}_{\mc{C}} \) at
the beginning of Stage~1.

\subsection{Main Analysis: Two-Stage Evolution and Feasible Bounds on $\Jxx$}

The main idea of the analysis is to exploit the \(\Jxx\)-induced see-saw reshaping
of the signed block spectrum, to remove the original (\LM{},\GM{})-anti-crossing.
While increasing \( \Jxx \) lifts the energies associated with \LM{} in the
same-sign block, it simultaneously lowers those of the opposite-sign blocks,
introducing a trade-off that constrains the allowable range of \( \Jxx \).
If \( \Jxx \) is too small, the local minima in the same-sign block are not
lifted sufficiently to remove the anti-crossing.
If \( \Jxx \) is too large, the opposite-sign blocks are lowered too far,
leading to new anti-crossings.
As a result, the success of the algorithm depends critically on choosing
\( \Jxx \) within an appropriate window.

We divide the evolution into two stages: Stage~1 and Stage~2.
Stage~2 addresses the original (\LM{},\GM{})-anti-crossing and removes it via
the see-saw effect; see Figure~\ref{fig:see-saw}(c) for an illustration.
Stage~1 ensures that the ground state evolves smoothly into the
\GM-support region---which serves as the starting point for Stage~2---without
encountering a new anti-crossing.
In terms of parameters, the \(\XX\)-coupling strength remains fixed at
\( \Jxx \) throughout Stage~1 and is reduced to zero during Stage~2.

We now briefly summarize the role of \( \Jxx \) bounds and how they jointly ensure a
smooth two-stage evolution.
The Stage-Separation bound \( \Jxxsep \) together with a \(\Jzz\) bound ensure that Stage~1 is effectively confined to the same-sign block, 
allowing Stage~1 and Stage~2 to be analyzed separately; 
the Lifting bound \( \Jxxlift \) ensures the original anti-crossing is removed; 
the Steering bound \( \Jxxsteer \) directs the system smoothly into the \(R\)-localized region (bypassing tunneling) during Stage~1,
and Stage~2 is secured by the Sinking bound  \( \Jxxsink \), which prevents the emergence of a new anti-crossing when the lowest opposite-sign block participates. 
The four analytical feasibility bounds on \( \Jxx \) are summarized in
Table~2 in Section~9.3 of Supplement.

The resulting evolution can be understood in terms of two key
mechanisms:
(i) \emph{structural steering} in Stage~1, which guides the evolving ground state
away from the \LM{}-supporting region and directly into the
\GM{}-supporting region; and
(ii) \emph{sign-generating quantum interference} in Stage~2, which enables smooth
adiabatic evolution to \GM{} via an opposite-sign path.

\subsubsection{Stage~1: Smooth Steering into the \GM-Support Region}
\label{sec:stage1}

The purpose of Stage~1 is to steer the ground state smoothly into the
\GM-support region through \emph{structural steering},
which then serves as the initial condition for Stage~2,
without encountering an anti-crossing.
Throughout this stage, the dynamics remains confined to the same-sign block,
ensured by the Stage-Separation bound \( \Jxxsep \), together with
the requirement \( \Jzz \le \Jzzinter \) in the shared-structure case.

\begin{figure}[!htbp]
  \centering
  \includegraphics[width=0.6\textwidth]{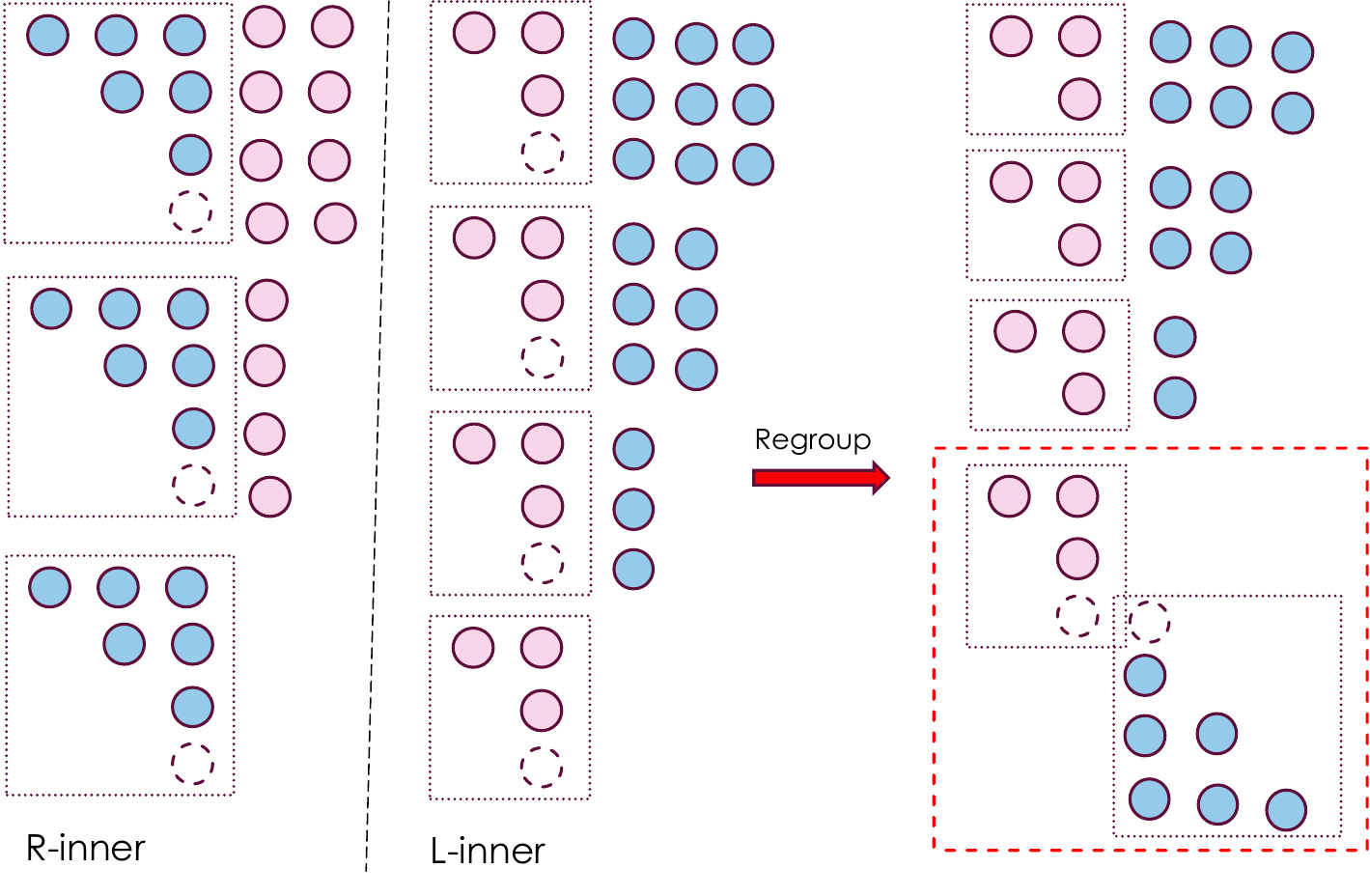}\\
  (\textbf{a}) \hspace*{4cm} (\textbf{b})\hspace*{3cm}
\caption{
\textbf{Two equivalent inner orderings of the basis states of the same-sign block
(symmetric subspace).} Illustrated for \( m = 2 \), \( m_r = 3 \).
(\textbf{a}) \emph{\(R\)-inner} ordering, in which basis states are grouped by the
spin-up count in the \(L\) subsystem (pink), with each group forming a block
over the \(R\) subsystem (blue). The dashed circle marks the empty-set basis state.
For appropriately large \( \Jxx \), the adiabatic evolution follows the
\(R\)-inner decomposition, leading to \(R\)-localization.
(\textbf{b}) \emph{\(L\)-inner} ordering, obtained by a permutation of basis states,
in which basis states are grouped by the spin-up count in the \(R\) subsystem.
For \( \Jxx = 0 \), the evolution follows the \(L\)-inner decomposition,
leading to \(L\)-localization.
The red dashed box highlights the core Hamiltonian \( \Hcore \), corresponding to
the reduced two-block structure resulting from \(L\)-localization.
}

\label{fig:L-R-inner}
\end{figure}

The structural steering mechanism can be understood through two equivalent
inner decompositions of the same-sign block: the 
\emph{\(L\)-inner} and the \emph{\(R\)-inner} ordering of basis states
(Fig.~\ref{fig:L-R-inner}).
Based on the inner decompositions, we can then describe the two different localizations.
If the evolution follows the \(R\)-inner (\(L\)-inner, resp.) decomposition,
the instantaneous ground state localizes within the lowest-energy
\(R\)-blocks, which we refer to as \emph{\(R\)-localization}
(\emph{\(L\)-localization}, resp.).

In the TFQA case (i.e.\ \( \Jxx = 0 \)), the evolution follows the \(L\)-inner decomposition,
leading to \(L\)-localization first.
The dynamics reduces to the two-block Hamiltonian
\( \Hcore \), where the subsequent tunneling transition from \(L\)-localized to
\(R\)-localized configurations proceeds through an \((L,R)\)-anti-crossing, as
analyzed in Section~11 of Supplement.

However, if \( \Jxx \) is appropriately large, the adiabatic evolution follows the
\(R\)-inner decomposition, leading to \(R\)-localization, without tunneling. 
Formally, the smooth \(R\)-localization is guaranteed by two
feasibility bounds:
the Lifting bound \(\Jxxlift\), which forces the ground state toward the lowest \(R\)-blocks;
and the Steering bound \(\Jxxsteer\), which ensures that this localization proceeds smoothly.  
This analytically established behavior is further confirmed by numerical evidence provided in
Section~9.3.2 of Supplement.

\subsubsection{Stage~2: Smooth Evolution to \GM{} via an Opposite-Sign Path}
\label{sec:stage2}

The purpose of Stage~2 is to remove the original (\LM{},\GM{})-anti-crossing and
complete the evolution to the global minimum (\GM{}) without a new anti-crossing,
starting from the
\(R\)-localized region prepared at the end of Stage~1.
This is achieved by exploiting the see-saw reshaping of the spectrum induced by
the \(\XX\)-coupling, together with quantum interference between opposite sign
sectors.

The value of \( \Jxx \) at the start of Stage~2 is chosen sufficiently large,
as ensured by the Lifting bound \(\Jxxlift\), so that the original anti-crossing is removed.
In the disjoint-structure graph \( \Gdis \), the blocks are decoupled, the evolution during Stage~2
remains entirely within the same-sign block. Thus there is no anti-crossing in this case.

In the shared-structure graph \( \Gshare \), the lowest opposite-sign block
(the AS0 block) enters the low-energy spectrum during Stage~2.
The Sinking upper bound \( \Jxx \le \Jxxsink \) guarantees that this
participation does not generate a new anti-crossing, so the evolution remains
smooth.
In terms of evolution, the involvement of opposite-sign sectors allows the ground state to
develop negative amplitudes in the computational basis.
These arise from constructive and destructive interference between same-sign
and opposite-sign components, a mechanism we refer to as
\emph{sign-generating quantum interference}.
This interference-enabled path to the global
minimum is inaccessible to stoquastic annealing, where amplitudes remain
strictly nonnegative.
The numerical confirmation of Stage~2 is given in Section~9.6 of Supplement.

Taken together, the two stages yield a smooth overall evolution:
Stage~1 steers the system into the \(R\)-localized region without tunneling,
and Stage~2 completes the transition to \GM{} via an opposite-sign,
interference-enabled path.

%4569

%%% %Methods
%% \input{Overview0123.tex} %1561
%% \input{MethodDecomp0123.tex}%1670
%% \input{MethodAnalysis0123.tex} %1115

%%%%%%%%%%%%%%%% REFERENCES %%%%%%%%%%%%%%%

%\clearpage % Clear all remaining figures and tables then start a new page

% The list of references goes after the main text and before the acknowledgements
% When preparing an initial submission, we recommend you use BibTeX, like this:
%
%\bibliography{Kabob0123} % for a file named science_template.bib

\begin{thebibliography}{10}
\providecommand{\url}[1]{\texttt{#1}}
\expandafter\ifx\csname urlstyle\endcsname\relax
  \providecommand{\doi}[1]{doi:\discretionary{}{}{}#1}\else
  \providecommand{\doi}{doi:\discretionary{}{}{}\begingroup
  \urlstyle{rm}\Url}\fi

\bibitem{Shor1997}
P.~W. Shor, Polynomial-Time Algorithms for Prime Factorization and Discrete
  Logarithms on a Quantum Computer. \emph{SIAM Journal on Computing}
  \textbf{26}~(5), 1484--1509 (1997).

\bibitem{Babbush2025GrandChallenge}
R.~Babbush, \emph{et~al.}, The Grand Challenge of Quantum Applications.
  \emph{arXiv preprint arXiv:2511.09124v3}  (2025).

\bibitem{Dupont2023SciAdv}
M.~Dupont, \emph{et~al.}, Quantum-enhanced greedy combinatorial optimization
  solver. \emph{Science Advances} \textbf{9}~(45), eadi0487 (2023).

\bibitem{Jordan2025DQI}
S.~P. Jordan, \emph{et~al.}, Optimization by Decoded Quantum Interferometry.
  \emph{Nature} \textbf{646}, 831--836 (2025).

\bibitem{Choi2021}
V.~Choi, Essentiality of the Non-stoquastic Hamiltonians and Driver Graph
  Design in Quantum Optimization Annealing. \emph{arXiv preprint
  arXiv:2105.02110v2}  (2021).

\bibitem{GareyJohnson1979}
M.~R. Garey, D.~S. Johnson, \emph{Computers and Intractability: A Guide to the
  Theory of NP-Completeness} (W. H. Freeman, San Francisco) (1979).

\bibitem{Choi2008}
V.~Choi, Minor-embedding in adiabatic quantum computation: I. The parameter
  setting problem. \emph{Quantum Information Processing} \textbf{7}~(5),
  193--209 (2008).

\bibitem{Choi2011}
V.~Choi, Minor-embedding in adiabatic quantum computation: II. Minor-universal
  graph design. \emph{Quantum Information Processing} \textbf{10}~(3), 343--353
  (2011).

\bibitem{Johnson2011}
M.~W. Johnson, \emph{et~al.}, Quantum annealing with manufactured spins.
  \emph{Nature} \textbf{473}~(7346), 194--198 (2011).

\bibitem{Ebadi2022ScienceMIS}
S.~Ebadi, \emph{et~al.}, Quantum optimization of maximum independent set using
  Rydberg atom arrays. \emph{Science} \textbf{376}~(6598), 1209--1215 (2022).

\bibitem{Farhi2000}
E.~Farhi, J.~Goldstone, S.~Gutmann, M.~Sipser, Quantum Computation by Adiabatic
  Evolution. \emph{arXiv preprint quant-ph/0001106}  (2000).

\bibitem{Farhi2001}
E.~Farhi, \emph{et~al.}, A Quantum Adiabatic Evolution Algorithm Applied to
  Random Instances of an NP-Complete Problem. \emph{Science}
  \textbf{292}~(5516), 472--475 (2001).

\bibitem{Nishimori2001SpinGlasses}
H.~Nishimori, \emph{Statistical Physics of Spin Glasses and Information
  Processing: An Introduction} (Oxford University Press, Oxford, UK) (2001).

\bibitem{Aharonov2007AQCEq}
D.~Aharonov, \emph{et~al.}, Adiabatic quantum computation is equivalent to
  standard quantum computation. \emph{SIAM Journal on Computing}
  \textbf{37}~(1), 166--194 (2007).

\bibitem{Barends2016NatureDAQC}
R.~Barends, \emph{et~al.}, Digitized adiabatic quantum computing with a
  superconducting circuit. \emph{Nature} \textbf{534}~(7606), 222--226 (2016).

\bibitem{AQC-Review}
T.~Albash, D.~A. Lidar, Adiabatic quantum computation. \emph{Rev. Mod. Phys.}
  \textbf{90}, 015002 (2018).

\bibitem{Beyond}
V.~Choi, Beyond Stoquasticity: Structural Steering and Interference in Quantum
  Optimization. \emph{arXiv preprint arXiv:2509.16263}  (2025).

\bibitem{Lamm2017ReduMIS}
S.~Lamm, P.~Sanders, C.~Schulz, D.~Strash, R.~F. Werneck, Finding Near-Optimal
  Independent Sets at Scale. \emph{Journal of Experimental Algorithmics}
  \textbf{23}, 1--20 (2018).

\bibitem{Lamm2016KaMIS}
S.~Lamm, P.~Sanders, C.~Schulz, D.~Strash, R.~F. Werneck, KaMIS: An Exact
  Solver for the Maximum Independent Set Problem, in \emph{Proceedings of the
  18th Workshop on Algorithm Engineering and Experiments (ALENEX)} (2016).

\bibitem{Hukushima1996PT}
K.~Hukushima, K.~Nemoto, Exchange Monte Carlo Method and Application to Spin
  Glass Simulations. \emph{Journal of the Physical Society of Japan}
  \textbf{65}~(6), 1604--1608 (1996).

\bibitem{KKR2006}
J.~Kempe, A.~Kitaev, O.~Regev, The Complexity of the Local Hamiltonian Problem.
  \emph{SIAM Journal on Computing} \textbf{35}~(5), 1070--1097 (2006).

\bibitem{Hen2021}
I.~Hen, Determining quantum Monte Carlo simulability with geometric phases.
  \emph{Phys. Rev. Research} \textbf{3}~(2), 023080 (2021).

\bibitem{Huang2025}
H.-Y. Huang, S.~Choi, J.~R. McClean, J.~Preskill, The Vast World of Quantum
  Advantage. \emph{arXiv:2508.05720}  (2025).

\bibitem{Farhi2014QAOA}
E.~Farhi, J.~Goldstone, S.~Gutmann, A Quantum Approximate Optimization
  Algorithm. \emph{arXiv:1411.4028}  (2014).

\bibitem{Limit}
V.~Choi, Limitation of Stoquastic Quantum Annealing: A Structural Perspective.
  \emph{arXiv preprint arXiv:2509.16265}  (2025).

\end{thebibliography}
%\bibliographystyle{sciencemag}

%%%%%%%%%%%%%%%% ACKNOWLEDGEMENTS %%%%%%%%%%%%%%%

\section*{Acknowledgments}
The author thanks Jamie Kerman for introducing her to the angular-momentum basis and for early discussions.
The author also thanks Siyuan Han for helpful comments.

The author acknowledges support from the Defense Advanced Research Projects Agency under Air Force Contract No. FA8702-15-D-0001. Any opinions, findings and conclusions or recommendations expressed in this material are those of the author and do not necessarily reflect the views of the Defense Advanced Research Projects Agency.

%% \paragraph*{Author contributions:}
%% V.C. conceived the project, designed the algorithm, developed the theoretical framework, performed all analyses, and wrote the manuscript.

%% \paragraph*{Competing interests:}
%% There are no competing interests to declare.

%% \paragraph*{Data and materials availability:}
%% All data needed to evaluate the conclusions in the paper are present in the paper and/or the Supplementary Information.

\subsection*{Supplementary materials}
Supplement consists of Sections~1--10, which correspond to
Sections~1--10 of \cite{Beyond}, and Section~11, which corresponds to Section~5
of \cite{Limit}.

\end{document}